\documentclass[12pt] {article}
\begin{document}
\setcounter{page}{1}
\def\theequation{\arabic{section}.\arabic{equation}}
\def\theequation{\thesection.\arabic{equation}}
\setcounter{section}{0}

\title{Neutron--proton radiative capture,\\ photo--magnetic  and 
anti--neutrino disintegration of the deuteron in the relativistic 
field theory model of the deuteron}

\author{A. N. Ivanov~\thanks{E--mail: ivanov@kph.tuwien.ac.at, Tel.:
+43--1--58801--14261, Fax: +43--1--58801--14299}~${\textstyle
^\ddagger}$, H. Oberhummer~\thanks{E--mail: ohu@kph.tuwien.ac.at,
Tel.: +43--1--58801--14251, Fax: +43--1--58801--14299} ,
N. I. Troitskaya~\thanks{Permanent Address: State Technical
University, Department of Nuclear Physics, 195251 St. Petersburg,
Russian Federation} , M. Faber~\thanks{E--mail:
faber@kph.tuwien.ac.at, Tel.: +43--1--58801--14261, Fax:
+43--1--58801--14299}}

\date{\today}

\maketitle

\begin{center}
{\it Institut f\"ur Kernphysik, Technische Universit\"at Wien,\\
Wiedner Hauptstr. 8-10, A-1040 Vienna, Austria}
\end{center}

\begin{center}
\begin{abstract}
The cross sections for the M1--capture n + p $\to$ D + $\gamma$, the
photo--magnetic and anti--neutrino disintegration of the deuteron are
evaluated in the relativistic field theory model of the deuteron
(RFMD). The cross section for M1--capture is evaluated by taking into
account the contributions of chiral one--meson loop corrections and
the $\Delta(1232)$ resonance. The cross sections for the
photo--magnetic and anti--neutrino disintegration of the deuteron are
evaluated by accounting for final--state interaction of the nucleon
pair in the ${^1}{\rm S}_0$--state. The amplitudes of low--energy
elastic np and nn scattering contributing to these processes are
obtained in terms of the S--wave scattering lengths and the effective
ranges. This relaxes substantially the statement by Bahcall and
Kamionkowski (Nucl. Phys. A625 (1997) 893) that the RFMD is unable to
describe a non--zero effective range for low--energy elastic
nucleon--nucleon scattering. The cross sections for the anti--neutrino
disintegration of the deuteron averaged over the anti--neutrino energy
spectrum agree good with experimental data.
\end{abstract}
\end{center}

\begin{center}
\end{center}

\newpage

\section{Introduction}
\setcounter{equation}{0}

\hspace{0.2in} As we have shown in Ref.\,[1] the relativistic field
theory model of the deuteron (RFMD) [2--6] is motivated by QCD. The
deuteron appears as a neutron--proton collective excitation -- a
Cooper np--pair induced by a phenomenological local four--nucleon
interaction in the nuclear phase of QCD. Strong low--energy
interactions of the deuteron coupled to itself and other particles are
described in terms of one--nucleon loop exchanges. The one--nucleon
loop exchanges allow to transfer nuclear flavours from an initial to a
final nuclear state by a minimal way and to take into account
contributions of nucleon--loop anomalies determined completely by
one--nucleon loop diagrams. The dominance of contributions of
nucleon--loop anomalies has been justified in the large $N_C$
expansion, where $N_C$ is the number of quark colours [1]\footnote{In
Ref.[6] we have considered a modified version of the RFMD which is not
well defined due to a violation of Lorentz invariance of the effective
four--nucleon interaction describing N + N $\to$ N + N
transitions. This violation has turned out to be incompatible with a
dominance of one--nucleon loop anomalies which are Lorentz
covariant.Thereby, the astrophysical factor for the solar proton
burning calculated in Ref.[5] and enhanced by a factor of 1.4 with
respect to the recommended value (E. G. Adelberger {\it et al.},
Rev. Mod. Phys. 70 (1998) 1265) is not good established. }.

In this paper we apply the RFMD to the evaluation of the cross
sections for the radiative M1--capture n + p $\to$ D + $\gamma$ for
thermal neutrons caused by the ${^1}{\rm S}_0 \to {^3}{\rm S}_1$
transition, the photo--magnetic $\gamma$ + D $\to$ n + p and
anti--neutrino disintegration of the deuteron caused by charged
$\bar{\nu}_{\rm e}$ + D $\to$ e$^+$ + n + n and neutral
$\bar{\nu}_{\rm e}$ + D $\to$ $\bar{\nu}_{\rm e}$ + n + p weak
currents.  We would like to emphasize that the main goal of the paper
is to show that: 1) Chiral perturbation theory can be incorporated in
the RFMD, and 2) the amplitudes of low--energy elastic
nucleon--nucleon scattering contributing to the reactions of the
photo--magnetic and anti--neutrino disintegration of the deuteron can
be described in the RFMD in agreement with low--energy nuclear
phenomenology.

As has been found in Refs.[4--6] the cross section for the M1--capture
calculated in the RFMD in the tree--meson approximation $\sigma^{\rm
np} = 276\,{\rm mb}$ differs from the experimental data $\sigma^{\rm
np}_{\exp} = (334.2\pm 0.5)\,{\rm mb}$ [7] by about 17$\%$ of the
experimental value. However, as has been shown in Refs.\,[8]
contributions of chiral meson--loop corrections play an important role
for the correct description of the process n + p $\to$ D +
$\gamma$. The evaluation of these corrections demands the use of
Chiral perturbation theory. 

For the evaluation of chiral one--meson loop corrections in the RFMD
we use Chiral perturbation theory at the quark level (CHPT)$_q$ with a
linear realization of chiral $U(3)\times U(3)$ symmetry developed in
Refs.[9,10]. The main chiral one--meson loop corrections are induced
by following virtual meson transitions $\pi \to a_1\,\gamma$, $a_1 \to
\pi\,\gamma$, $\pi \to (\omega,\rho) \gamma$, $(\omega,\rho) \to \pi
\gamma$, $\sigma \to (\omega,\rho) \gamma$ and $(\omega,\rho) \to
\sigma \gamma$, where $\sigma$ is a scalar partner of pions under
chiral $SU(2)\times SU(2)$ symmetry [9,10].

However, as has been stated by Riska and Brown [11] (see also [17])
 for correct description of the amplitude of the neutron--proton
 radiative capture one needs to take into account the contribution of
 the $\Delta(1232)$ resonance. In the RFMD the contribution of the
 $\Delta(1232)$ resonance has been considered in Ref.[12] by example
 of the evaluation of the S--wave scattering length of low--energy
 elastic $\pi$D scattering.

At threshold the amplitude of the photo--magnetic disintegration of
the deuteron is related to the amplitude of the M1--capture. In order
to evaluate the amplitude of $\gamma$ + D $\to$ n + p for the energy
region far from threshold we take into account the contributions of
the np interaction in the final state. For this aim we sum up an
infinite series of one--nucleon loop diagrams and evaluate the result
of the summation in leading order in the large $N_C$ expansion
[1]. This gives the amplitude of low--energy elastic np scattering
contributing to the amplitude of $\gamma$ + D $\to$ n + p defined by
the S--wave scattering length and the effective range.

The developed technique we apply to the evaluation of the cross
section for the anti--neutrino disintegration of the deuteron
$\bar{\nu}_{\rm e}$ + D $\to$ e$^+$ + n + n and $\bar{\nu}_{\rm e}$ +
D $\to$ $\bar{\nu}_{\rm e}$ + n + p. The reaction $\bar{\nu}_{\rm e}$
+ D $\to$ e$^+$ + n + n is caused by charged weak current and valued,
in the sense of charge independence of weak interaction strength, to
be equivalent to the observation of the reaction of the solar proton
burning, or pp fusion, p + p $\to$ D + e$^+$ + $\nu_{\rm e}$ in the
terrestrial laboratories [13]. We compare the theoretical cross
sections with recent experimental data given by the Reines's
experimental group [14].

For the description of low--energy transitions N + N $\to$ N + N in
the reactions n + p $\to$ D + $\gamma$, $\gamma$ + D $\to$ n + p,
$\bar{\nu}_{\rm e}$ + D $\to$ e$^+$ + n + n, $\bar{\nu}_{\rm e}$ +
D $\to$ $\bar{\nu}_{\rm e}$ + n + p  and p + p $\to$ D + e$^+$
+ $\nu_{\rm e}$, where nucleons are in the ${^1}{\rm S}_0$--state, we
apply the effective local four--nucleon interactions [2--5]:
\begin{eqnarray}\label{label1.1}
&&{\cal L}^{\rm NN \to NN}_{\rm eff}(x)=G_{\rm \pi
NN}\,\{[\bar{n}(x)\gamma_{\mu}
\gamma^5 p^c(x)][\bar{p^c}(x)\gamma^{\mu}\gamma^5 n(x)]\nonumber\\
&&+\frac{1}{2}\,
[\bar{n}(x)\gamma_{\mu} \gamma^5 n^c(x)][\bar{n^c}(x)\gamma^{\mu}\gamma^5
n(x)] +
\frac{1}{2}\,[\bar{p}(x)\gamma_{\mu} \gamma^5 p^c(x)]
[\bar{p^c}(x)\gamma^{\mu}\gamma^5 p(x)]\nonumber\\
&&+ (\gamma_{\mu}\gamma^5 \otimes \gamma^{\mu}\gamma^5 \to \gamma^5 \otimes
\gamma^5)\},
\end{eqnarray}
where $n(x)$ and $p(x)$ are the operators of the neutron and the proton
interpolating fields, $n^c(x) = C \bar{n}^T(x)$ and so on, then $C$ is a charge
conjugation matrix and $T$ is a transposition.
The effective coupling constant $G_{\rm \pi NN}$ is defined by [3--5]
\begin{eqnarray}\label{label1.2}
G_{\rm \pi NN} = \frac{g^2_{\rm \pi NN}}{4M^2_{\pi}} - \frac{2\pi a_{\rm
np}}{M_{\rm N}} = 3.27\times 10^{-3}\,{\rm MeV}^{-2},
\end{eqnarray}
where $g_{\rm \pi NN}= 13.4$ is the coupling constant of the ${\rm \pi
NN}$ interaction, $M_{\pi}=135\,{\rm MeV}$ is the pion mass, $M_{\rm
p} = M_{\rm n} = M_{\rm N} = 940\,{\rm MeV}$ is the mass of the proton
and the neutron neglecting the electromagnetic mass difference, which
is taken into account only for the calculation of the phase volumes of
the final states of the reactions $\bar{\nu}_{\rm e}$ + D $\to$ e$^+$
+ n + n, $\bar{\nu}_{\rm e}$ + D $\to$ $\bar{\nu}_{\rm e}$ + n + p and
p + p $\to$ D + e$^+$ + $\nu_{\rm e}$, and $a_{\rm np} = (-23.75\pm
0.01)\,{\rm fm}$ is the S--wave scattering length of np scattering in
the ${^1}{\rm S}_0$--state.

The first term in the effective coupling constant $G_{\rm \pi NN}$
comes from the one--pion exchange for the squared momenta transfer $-
q^2$ much less than the squared pion mass $- q^2 \ll M^2_{\pi}$ and
the subsequent Fierz transformation of nucleon fields (see
Appendix B of Ref.\,[6]). We should emphasize that due to Fierz
transformation the effective local four--nucleon interaction caused by
the one--pion exchange contains a few contributions with different
spinorial structure, we have taken into account only those terms which
contribute to the ${^1}{\rm S}_0$--state of the NN system. The second
term in Eq.(\ref{label1.2}) is a phenomenological one representing a
collective contribution caused by the integration over heavy meson
degrees of freedom [5,6]. This term is taken in the form used in the
Effective Field Theory (EFT) approach [15,16] for the description of
low--energy elastic NN scattering. The effective interaction
Eq.\,(\ref{label1.1}) is written in isotopically invariant form,
and the coupling constant $G_{\rm \pi NN}$ can be never equal to zero
at $a_{\rm np} \neq 0$ due to a negative value of $a_{\rm np}$ imposed
by nuclear forces, $a_{\rm np} < 0$ [17]. Note that the contribution
of the phenomenological part to the effective coupling constant
$G_{\rm \pi NN}$ makes up less than 33$\%$.

In the low--energy limit the effective local four--nucleon interaction
Eq.\,(\ref{label1.1}) vanishes due to the reduction
\begin{eqnarray}\label{label1.3}
[\bar{N}(x)\gamma_{\mu}\gamma^5 N^c (x)][\bar{N^c}(x)\gamma^{\mu}\gamma^5
N(x)] \to - [\bar{N}(x) \gamma^5 N^c(x)][\bar{N^c}(x) \gamma^5 N(x)],
\end{eqnarray}
where $N(x)$ is the neutron or the proton interpolating field. Such a
vanishing of the one--pion exchange contribution to the NN potential
is well--known in the EFT approach [15,16] and the potential model
approach (PMA) [17]. In power counting [15,16] the interaction induced
by the one--pion exchange is of order $O(k^2)$, where $k$ is a
relative momentum of the NN system. The former is due the Dirac matrix
$\gamma^5$ which leads to the interaction between small components of
Dirac bispinors of nucleon wave functions.

In the one--nucleon loop exchange approach the contributions of the
interactions $[\bar{N}(x)\gamma_{\mu}\gamma^5 N^c
(x)][\bar{N^c}(x)\gamma^{\mu}\gamma^5 N(x)]$ and $[\bar{N}(x) \gamma^5
N^c(x)][\bar{N^c}(x) \gamma^5 N(x)]$ to amplitudes of nuclear
processes are different and do not cancel each other in the
low--energy limit due to the dominance of nucleon--loop anomalies
[1]. This provides the interaction between large components of Dirac
bispinors of nucleon wave functions.

The paper is organized as follows. In Sect.\,2 we evaluate the
contribution of chiral one--meson loop corrections to the amplitude of
the neutron--proton radiative capture and the cross section for the
neutron--proton radiative capture.  In Sect.\,3 we include the
contribution of the $\Delta(1232)$ resonance and analyse the total
cross section for the neutron--proton radiative capture for thermal
neutrons and compare it with experimental data.  In Sect.\,4 we
evaluate the cross section for $\gamma$ + D $\to$ n + p for energies
far from threshold. The contribution of low--energy elastic np
scattering to the amplitude of the process $\gamma$ + D $\to$ n + p is
evaluated in agreement with low--energy nuclear phenomenology.  This
relaxes substantially the statement by Bahcall and Kamionkowski [18]
that in the RFMD due to the local four--nucleon interaction
Eq.(\ref{label1.1}) one cannot describe low--energy elastic NN
scattering in agreement with low--energy nuclear
phenomenology. However, the problem of the description of low--energy
elastic pp scattering accounting for the Coulomb repulsion still
remains. In Sects.\,5 and 6 we evaluate the cross sections for the
anti--neutrino disintegration of the deuteron caused by charged
$\bar{\nu}_{\rm e}$ + D $\to$ e$^+$ + n + n and neutral
$\bar{\nu}_{\rm e}$ + D $\to$ $\bar{\nu}_{\rm e}$ + n + p weak
currents and average them over the anti--neutrino energy spectrum. The
average values of the cross sections agree good with experimental
data. In the Conclusion we discuss the obtained results.

\section{Neutron--proton radiative capture}
\setcounter{equation}{0}

\hspace{0.2in} At low energies the neutron--proton radiative capture n
+ p $\to$ D + $\gamma$ runs through the magnetic dipole transition
${^1}{\rm S}_0 \to {^3}{\rm S}_1$, the M1--capture.  In the RFMD the
amplitude of the M1--capture calculated in the tree--meson
approximation reads [3--6]\footnote{For the details of the calculation
we relegate readers to Appendix F of Ref.\,[6].}
\begin{eqnarray}\label{label2.1}
{\cal M}({\rm n + p} \to {\rm D + \gamma}) =\,e\,(\mu_{\rm p} -
\mu_{\rm n})\,\frac{5 g_{\rm V}}{8\pi^2}\,G_{\rm \pi
NN}\,\,(\vec{q}\times \vec{e}^{\,*}(\vec{q}\,))\cdot
\vec{e}^{\,*}(\vec{k}_{\rm D}) \,[\bar{u^c}(p_2)\gamma^5 u(p_1)],
\end{eqnarray}
where $e$ is the proton electric charge, $\mu_{\rm p} = 2.793$ and
$\mu_{\rm n} = - 1.913$ are the magnetic dipole moments of the proton
and the neutron, respectively, measured in nuclear magnetons, $g_{\rm
V}$ is a phenomenological coupling constant of the RFMD related to the
electric quadrupole moment of the deuteron $Q_{\rm D} = 0.286\,{\rm
fm}^2$ [3]: $g^2_{\rm V} = 2\pi^2 Q_{\rm D}M^2_{\rm N}$; $\vec{q}$ and
$\vec{k}_{\rm D}$ are 3--momenta of the photon and the deuteron, and
$\vec{e}^{\,*}(\vec{q}\,)$ and $\vec{e}^{\,*}(\vec{k}_{\rm D})$ -- the
polarization vectors of them; $\bar{u^c}(p_2) $ and $u(p_1)$ are the
Dirac bispinors of the neutron and the proton.

The cross section for the M1--capture calculated in the tree--meson
approximation is then defined [3--6]: 
\begin{eqnarray}\label{label2.2}
\sigma^{\rm n p}(k) = \frac{1}{v}\,(\mu_{\rm p}-\mu_{\rm
n})^2\,\frac{25}{64}\,\frac{\alpha}{\pi^2}\,\,Q_{\rm D}\,G^2_{\rm \pi
NN}\,M_{\rm N}\varepsilon^3_{\rm D} = 276\,{\rm m b},
\end{eqnarray}
where $k$ is a relative momentum of the np system.  The numerical
value has been computed for $k = 0$, $\epsilon_{\rm D} = 2.225\,{\rm
MeV}$ and $v = 7.34\,\times 10^{-6}$ (the absolute value $v =
2.2\,\times 10^5\,{\rm cm}\,{\rm s}^{-1}$), the laboratory velocity of
the neutron. The theoretical value $\sigma^{\rm n p}(k) = 276\,{\rm m b}$
agrees within an accuracy better than 10$\%$ with the theoretical
value [17]
\begin{eqnarray}\label{label2.3}
\sigma^{\rm n p}_{\rm PMA}(k) = (302.5\pm 4)\,{\rm mb}
\end{eqnarray}
calculated in the PMA for the pure M1 transition. In comparison with
the experimental data [7]
\begin{eqnarray}\label{label2.4}
\sigma^{\rm n p}_{\exp}\,=\,(334.2\pm 0.5)\,{\rm
mb}
\end{eqnarray}
the theoretical value Eq.(\ref{label2.2}) obtained in the RFMD is less
by 17$\%$ of the experimental one.

However, as has been shown in Refs.\,[8] chiral meson--loop
corrections play an important role for the correct description of the
low--energy process n + p $\to$ D + $\gamma$ for thermal neutrons. The
evaluation of chiral meson--loop corrections in the RFMD we use
(CHPT)$_q$ developed in Refs.[9,10]. Below we consider the
contributions of chiral one--meson loop corrections induced by the
virtual meson transitions $\pi \to a_1 \gamma$, $a_1 \to \pi\,\gamma$,
$\pi \to (\omega, \rho) \gamma$, $(\omega, \rho) \to \pi \gamma$,
$\sigma \to (\omega, \rho)\gamma$ and $(\omega, \rho) \to \sigma
\gamma$, where $\sigma$ is a scalar partner of pions under chiral
$SU(2)\times SU(2)$ transformations in (CHPT)$_q$ with a linear
realization of chiral $U(3)\times U(3)$ symmetry [9,10].

The effective Lagrangians $\delta {\cal L}^{\rm pp\gamma}_{\rm
eff}(x)$ and $\delta {\cal L}^{\rm nn\gamma}_{\rm eff}(x)$, caused by
the virtual meson transitions $\pi \to a_1\,\gamma$, $a_1 \to
\pi\,\gamma$, $\pi \to (\omega, \rho) \gamma$, $(\omega, \rho) \to
\pi \gamma$, $\sigma \to (\omega, \rho)\gamma$ and $(\omega, \rho) \to
\sigma \gamma$, we evaluate in leading order in the large $N_C$
expansion [1]. The results of the evaluation contain divergent
contributions. In order to remove these divergences we apply the
renormalization procedure developed in (CHPT)$_q$ for the evaluation
of chiral meson--loop corrections (see {\it Ivanov} in
Refs. [9]). Since the renormalized expressions should vanish in the
chiral limit $M_{\pi} \to 0$ [9], only the virtual meson transitions
with intermediate $\pi$--meson give non--trivial contributions. The
contributions of the virtual meson transitions with intermediate
$\sigma$--meson are found finite in the chiral limit and subtracted
according to the renormalization procedure [9]. Such a cancellation of
the $\sigma$--meson contributions in the one--meson loop approximation
agrees with Chiral perturbation theory using a non--linear realization
of chiral symmetry, where $\sigma$--meson like exchanges can appear
only in two--meson loop approximation.  Then, the sum of the
contributions of the virtual meson transitions $\pi^- \to \rho^-
\gamma$, $\pi^0 \to \rho^0 \gamma$ and $\pi^0 \to \omega \gamma$ to
the effective coupling ${\rm nn\gamma}$ is equal to zero. As a result
the effective Lagrangians $\delta {\cal L}^{\rm pp\gamma}_{\rm
eff}(x)$ and $\delta {\cal L}^{\rm nn\gamma}_{\rm eff}(x)$ are given
by
\begin{eqnarray}\label{label2.5}
\delta {\cal L}^{\rm pp\gamma}_{\rm eff}(x)&=& \frac{ie}{4M_{\rm
N}}\Bigg[g_{\rm A}g_{\rm \pi
NN}\frac{\alpha_{\rho}}{16\pi^3}\,\frac{M_{\rm
N}}{F_{\pi}}M^2_{\pi}\,J_{\rm \pi a_1 N } + g_{\rm \pi
NN}\,\frac{N_C\alpha_{\rho}}{16\pi^3}\frac{M_{\rm N}}{F_{\pi}}
\,M^2_{\pi}\,J_{\rm \pi VN}\Bigg]\,\nonumber\\
&&\times\,
[\bar{p}(x)\sigma_{\mu\nu}p(x)]\,F^{\mu\nu}(x),\nonumber\\ \delta
{\cal L}^{\rm nn\gamma}_{\rm eff}(x)&=&\frac{ie}{4M_{\rm
N}}\Bigg[-\,g_{\rm A}g_{\rm \pi
NN}\frac{\alpha_{\rho}}{16\pi^3}\,\frac{M_{\rm
N}}{F_{\pi}}M^2_{\pi}\,J_{\rm \pi a_1 N }\Bigg]\, 
[\bar{n}(x)\sigma_{\mu\nu}n(x)]\,F^{\mu\nu}(x),
\end{eqnarray}
where $F_{\mu\nu}(x) = \partial_{\mu} A_{\nu}(x) - \partial_{\nu}
A_{\mu}(x)$ is the electromagnetic field strength, $\alpha_{\rho} =
g^2_{\rho}/4\pi = 2.91$ is the effective coupling constant of the
$\rho \to \pi \pi$ decay, $F_{\pi} = 92.4\,{\rm MeV}$ is the leptonic
coupling constant of pions, and $g_{\rm A} =1.267$ [19].  Then,
$J_{\rm \pi a_1 N}$ and$J_{\rm \pi VN}$ are the momentum integrals
determined by
\begin{eqnarray}\label{label2.6}
J_{\rm \pi a_1  N}&=&\int \frac{d^4p}{\pi^2}\,\frac{1}{(M^2_{\pi} +
p^2)(M^2_{a_1} + p^2)(M^2_{\rm N} + p^2)} =
0.017\,M^{-2}_{\pi},\nonumber\\ J_{\rm \pi VN} &=&\int
\frac{d^4p}{\pi^2}\,\frac{1}{(M^2_{\pi} + p^2)(M^2_{\rm V} +
p^2)(M^2_{\rm N} + p^2)}= 0.024\,M^{-2}_{\pi},
\end{eqnarray}
where $p$ is Euclidean 4--momentum, $M_{\rm V} = M_{\rho} = M_{\omega}
= 770\,{\rm MeV}$ [19] and $M_{a_1} =\sqrt{2}\,M_{\rho}$ [9].

At $N_C = 3$ the cross section for the M1--capture accounting for the
contribution of the effective interaction Eq.(\ref{label2.5}) amounts
to
\begin{eqnarray}\label{label2.7}
\hspace{-0.5in}&&\sigma^{\rm n p}(k)=
\frac{1}{v}\,(\mu_{\rm p}-\mu_{\rm
n})^2\,\frac{25}{64}\,\frac{\alpha}{\pi^2}\,\,Q_{\rm D}\,G^2_{\rm \pi
NN}\,M_{\rm N}\,\varepsilon^3_{\rm D}\,\nonumber\\
\hspace{-0.5in}&&\times\,\Bigg[1 + \frac{g^2_{\rm \pi NN}}{\mu_{\rm p}
- \mu_{\rm
n}}\,\frac{M^2_{\pi}}{8\pi^2}\,\frac{\alpha_{\rho}}{\pi}\Bigg(J_{\rm
\pi a_1 N} + \frac{3}{2 g_{\rm A}}\,J_{\rm \pi VN}\Bigg)\Bigg]^2 =
287.2\,{\rm m b},
\end{eqnarray}
where we have used the relation $g_{\rm \pi NN} \simeq g_{\rm
A}\,M_{\rm N}/F_{\pi}$. The theoretical
value of the cross section for the neutron--proton radiative capture
given by Eq.(\ref{label2.7}) differs from the experimental one by
about 14$\%$. This discrepancy we describe by taking into account the
contribution of the $\Delta(1232)$ resonance.

\section{$\Delta(1232)$ resonance}
\setcounter{equation}{0}

In our consideration the $\Delta(1232)$ resonance is the Rarita--Schwinger
field [20] $\Delta^a_{\mu}(x)$, the isotopical index $a$ runs over $a =
1,2,3$, having the following free Lagrangian [21,22]:
\begin{eqnarray}\label{label3.1}
\hspace{-0.5in} {\cal L}^{\Delta}_{\rm kin}(x) =
\bar{\Delta}^a_{\mu}(x) [-(i\gamma^{\alpha}\partial_{\alpha} -
M_{\Delta}) \,g^{\mu\nu} +
\frac{1}{4}\gamma^{\mu}\gamma^{\beta}(i\gamma^{\alpha}\partial_{\alpha}
- M_{\Delta}) \gamma_{\beta}\gamma^{\nu}] \Delta^a_{\nu}(x),
\end{eqnarray}
where $M_{\Delta} = 1232\,{\rm MeV}$ is the mass of the $\Delta$
resonance field $\Delta^a_{\mu}(x)$. In terms of the eigenstates of
the electric charge operator the fields $\Delta^a_{\mu}(x)$ are given
by [12,21,22]
\begin{eqnarray}\label{label3.2}
\begin{array}{llcl}
&&\Delta^1_{\mu}(x) = \frac{1}{\sqrt{2}}\Biggr(\begin{array}{c}
\Delta^{++}_{\mu}(x) - \Delta^0_{\mu}(x)/\sqrt{3} \\
\Delta^+_{\mu}(x)/\sqrt{3} - \Delta^-_{\mu}(x)
\end{array}\Biggl)\,,\,
\Delta^2_{\mu}(x) = \frac{i}{\sqrt{2}}\Biggr(\begin{array}{c}
\Delta^{++}_{\mu}(x) + \Delta^0_{\mu}(x)/\sqrt{3} \\
\Delta^+_{\mu}(x)/\sqrt{3} + \Delta^-_{\mu}(x)
\end{array}\Biggl)\,,\\
&&\Delta^3_{\mu}(x) = -\sqrt{\frac{2}{3}}\Biggr(\begin{array}{c}
\Delta^+_{\mu}(x) \\ \Delta^0_{\mu}(x) \end{array}\Biggl).
\end{array}
\end{eqnarray}
The fields $\Delta^a_{\mu}(x)$ obey the subsidiary constraints:
$\partial^{\mu}\Delta^a_{\mu}(x) = \gamma^{\mu}\Delta^a_{\mu}(x) = 0$
[20--22]. The Green function of the free $\Delta$--field is determined
\begin{eqnarray}\label{label3.3}
\hspace{-0.5in}<0|{\rm T}(\Delta_{\mu}(x_1)\bar{\Delta}_{\nu}(x_2))|0>
= - i S_{\mu\nu}(x_1 - x_2).
\end{eqnarray}
In the momentum representation $S_{\mu\nu}(x)$ reads [12,21,22]:
\begin{eqnarray}\label{label3.4}
\hspace{-0.5in}S_{\mu\nu}(p) = \frac{1}{M_{\Delta} - \hat{p}}\Bigg( -
g_{\mu\nu} + \frac{1}{3}\gamma_{\mu}\gamma_{\nu} +
\frac{1}{3}\frac{\gamma_{\mu}p_{\nu} -
\gamma_{\nu}p_{\mu}}{M_{\Delta}} +
\frac{2}{3}\frac{p_{\mu}p_{\nu}}{M^2_{\Delta}}\Bigg).
\end{eqnarray}
The most general form of the ${\rm \pi N \Delta}$-- interaction
compatible with the requirements of chiral symmetry reads [21]:
\begin{eqnarray}\label{label3.5}
\hspace{-0.5in}&&{\cal L}_{\rm \pi N \Delta}(x) = \frac{g_{\rm \pi
N\Delta}}{2M_{\rm
N}}\bar{\Delta}^a_{\omega}(x)\Theta^{\omega\varphi}N(x)
\partial_{\varphi}\pi^a(x)
+ {\rm h.c.} = \nonumber\\
\hspace{-0.5in}&&= \frac{g_{\rm \pi N\Delta}}{\sqrt{6}M_{\rm
N}}\Bigg[\frac{1}{\sqrt{2}}\bar{\Delta}^+_{\omega}(x)\Theta^{\omega\varphi}
n(x) \partial_{\varphi}\pi^+(x) -
\frac{1}{\sqrt{2}}\bar{\Delta}^0_{\omega}(x)\Theta^{\omega\varphi}
p(x) \partial_{\varphi}\pi^-(x)\nonumber\\
\hspace{-0.5in}&&- \bar{\Delta}^+_{\omega}(x)\Theta^{\omega\varphi}
p(x) \partial_{\varphi}\pi^0(x) -
\bar{\Delta}^0_{\omega}(x)\Theta^{\omega\varphi} p(x)
\partial_{\varphi}\pi^0(x) + \ldots \Bigg].
\end{eqnarray}
The nucleon field $N(x)$ is the isotopical doublet with the components
$N(x) = (p(x), n(x))$, and $\pi^a(x)$ is the pion field with the
components $\pi^1(x) = (\pi^-(x) + \pi^+(x))/\sqrt{2}$, $\pi^2(x) =
(\pi^-(x) - \pi^+(x))/i\sqrt{2}$ and $\pi^3(x) = \pi^0(x)$. The tensor
$\Theta^{\omega\varphi}$ is given in Ref.\,[21]:
$\Theta^{\omega\varphi} = g^{\omega\varphi} - (Z +
1/2)\gamma^{\omega}\gamma^{\varphi}$, where the parameter $Z$ is
arbitrary. There is no consensus on the exact value of $Z$. From
theoretical point of view $Z=1/2$ is preferred [21].  Phenomenological
studies give only the bound $|Z| \le 1/2$ [23]. The empirical value of
the coupling constant $g_{\rm \pi N\Delta}$ relative to the coupling
constant $g_{\rm \pi NN}$ is $g_{\rm \pi N\Delta} = 2.12\,g_{\rm \pi
NN}$ [24].

Assuming that the transition $\Delta \to {\rm N} + \gamma$ is
primarily a magnetic one the effective Lagrangian describing the
$\Delta \to {\rm N} + \gamma$ decays can be determined as [25,26]:
\begin{eqnarray}\label{label3.6}
\hspace{-0.5in}&&{\cal L}_{\rm \gamma N \Delta}(x) = i e \frac{g_{\rm
\gamma N\Delta}}{2M_{\rm N}}\bar{N}(x) \gamma_{\alpha}\gamma^5
\Delta^3_{\beta}(x) F^{\beta\alpha}(x) + {\rm h.c.} = \nonumber\\
\hspace{-0.5in}&&= - \frac{ie}{\sqrt{6}}\frac{g_{\rm \pi
N\Delta}}{M_{\rm N}}[\bar{p}(x)\gamma_{\alpha}\gamma^5
\Delta^+_{\beta}(x) + \bar{n}(x)\gamma_{\alpha}\gamma^5
\Delta^0_{\beta}(x)]\,F^{\beta\alpha}(x) + {\rm h.c.},
\end{eqnarray}
where $F^{\alpha\beta}(x) = \partial^{\alpha}A^{\beta}(x) -
\partial^{\beta}A^{\alpha}(x)$ and $A^{\alpha}(x)$ is the operator of
the photon field. The empirical value of the coupling constant $g_{\rm
\gamma N\Delta}$ relative to the coupling constant $g_{\rm \pi NN}$ is
$g_{\rm \gamma N\Delta} = 0.32\,g_{\rm \pi NN}$ [27].

For the calculation of the amplitude of the neutron--proton radiative
capture in the RFMD we have to calculate the effective Lagrangian
describing the n + p $\to$ $\Delta$ + N transitions. Following the
general procedure expounded in Ref.\,[3] we obtain:
\begin{eqnarray}\label{label3.7}
\hspace{-0.5in}&&{\cal L}^{\rm np \to \Delta N}_{\rm eff}(x) = -
\frac{i}{\sqrt{6}}\frac{g_{\rm \pi N\Delta}}{M_{\rm N}}\frac{g_{\rm
\pi NN}}{4 M^2_{\pi}}\int d^4z\,\frac{\partial}{\partial
z_{\varphi}}\delta^{(4)}(z-x)\,\{[\bar{\Delta}^+_{\omega}(z)\,
{\Theta^{\omega}}_{\varphi}\, n^c(x)]\,\nonumber\\
\hspace{-0.5in}&& \times\,[\bar{n^c}(z)\gamma^5 p(x) +
\bar{n^c}(x)\gamma^5 p(z)]
-[\bar{\Delta}^0_{\omega}(z)\,{\Theta^{\omega}}_{\varphi}\,
p^c(x)]\,[\bar{n^c}(z)\gamma^5 p(x) +
\bar{n^c}(x)\gamma^5 p(z)]\nonumber\\
\hspace{-0.5in}&& + 1 \otimes \gamma^5 \to -\gamma_{\nu} \otimes
\gamma^{\nu}\gamma^5\}.
\end{eqnarray}
Using then the phenomenological Lagrangian
\begin{eqnarray}\label{label3.8}
{\cal L}_{\rm npD}(x) = -ig_{\rm V}[\bar{p^c}(x)\gamma^{\mu}n(x) -
\bar{n^c}(x)\gamma^{\mu}p(x)] D^{\dagger}_{\mu}(x)
\end{eqnarray}
the effective Lagrangian describing the contribution of the $\Delta$
resonance to the amplitude of the transition n + p $\to$ D + $\gamma$
is defined [5]
\begin{eqnarray}\label{label3.9}
\hspace{-0.5in}&&\int d^4x\,{\cal L}_{\rm np \to \Delta N \to
D\gamma}(x) = - \int d^4x_1 d^4x_2 d^4x_3\,<{\rm T}({\cal L}^{\rm np
\to \Delta N}_{\rm eff}(x_1){\cal L}_{\rm npD}(x_2){\cal L}_{\rm
\gamma N\Delta}(x_3))> = \nonumber\\
\hspace{-0.5in}&&= - \frac{i}{6}\frac{eg_{\rm V}}{M^2_{\rm
N}}\frac{g_{\rm \pi N\Delta}}{g_{\rm \pi NN}}\frac{g_{\rm \gamma
N\Delta}}{g_{\rm \pi NN}}\frac{g^3_{\rm \pi NN}}{4M^2_{\pi}}\int d^4x_1 d^4x_2 d^4x_3\int d^4z\,\frac{\partial}{\partial
z_{\varphi}}\delta^{(4)}(z-x_1)\,\nonumber\\
\hspace{-0.5in}&&\times\,{\rm T}([\bar{p^c}(x_1)\gamma^5 n(z) +
\bar{p^c}(z)\gamma^5
n(x_1)]\,D^{\dagger}_{\mu}(x_2)
F^{\alpha\beta}(x_3))\nonumber\\
\hspace{-0.5in}&&\times\Big\{<0|{\rm
T}([\bar{\Delta}^+_{\omega}(z)\,
{\Theta^{\omega}}_{\varphi}\,n^c(x_1)][\bar{p^c}(x_2)\gamma^{\mu}n(x_2) -
\bar{n^c}(x_2)\gamma^{\mu}p(x_2)][\bar{p}(x_3)\gamma_{\beta}\gamma^5
\Delta^+_{\alpha}(x_3)])|0>\nonumber\\
\hspace{-0.5in}&& - <0|{\rm
T}([\bar{\Delta}^0_{\omega}(z)\,{\Theta^{\omega}}_{\varphi}\,
p^c(x)][\bar{p^c}(x_2)\gamma^{\mu}n(x_2) -
\bar{n^c}(x_2)\gamma^{\mu}p(x_2)]\nonumber\\
\hspace{-0.5in}&&\times\,[\bar{n}(x_3)\gamma_{\beta}\gamma^5
\Delta^0_{\alpha}(x_3)])|0> + (\gamma^5 \otimes 1 \to -
\gamma_{\nu}\gamma^5 \otimes \gamma^{\nu})\Big\} =\nonumber\\
\hspace{-0.5in}&&= \frac{i}{3}\frac{eg_{\rm V}}{M_{\rm N}}\frac{g_{\rm
\pi N\Delta}}{g_{\rm \pi NN}}\frac{g_{\rm \gamma N\Delta}}{g_{\rm \pi
NN}}\frac{g^3_{\rm \pi NN}}{4M^2_{\pi}}\int d^4x_1
d^4x_2 d^4x_3\int
d^4z\,\frac{\partial}{\partial z_{\varphi}}\delta^{(4)}(z-x_1)\nonumber\\
\hspace{-0.5in}&&\times\,{\rm T}([\bar{p^c}(x_1)\gamma^5 n(z) + 
\bar{p^c}(z)\gamma^5 n(x_1)]\,D^{\dagger}_{\mu}(x_2)
F^{\alpha\beta}(x_3))\nonumber\\
\hspace{-0.5in}&&\times\,\Big\{<0|{\rm
T}([\bar{\Delta}^+_{\omega}(z)\,
{\Theta^{\omega}}_{\varphi}\, n^c(x_1)][\bar{n^c}(x_2)\gamma^{\mu}p(x_2)]
[\bar{p}(x_3)\gamma_{\beta}\gamma^5
\Delta^+_{\alpha}(x_3)])|0>\nonumber\\
\hspace{-0.5in}&& + <0|{\rm
T}([\bar{\Delta}^0_{\omega}(z)\,{\Theta^{\omega}}_{\varphi}\, p^c(x_1)]
[\bar{p^c}(x_2)\gamma^{\mu}n(x_2)] [\bar{n}(x_3)\gamma_{\beta}\gamma^5
\Delta^0_{\alpha}(x_3)])|0> \nonumber\\
\hspace{-0.5in}&& + (\gamma^5 \otimes 1 \to - \gamma_{\nu}\gamma^5
\otimes \gamma^{\nu})\Big\} =\nonumber\\
\hspace{-0.5in}&&= \frac{2}{3}\frac{ieg_{\rm V}}{M^2_{\rm
N}}\frac{g_{\rm \pi N\Delta}}{g_{\rm \pi NN}}\frac{g_{\rm \gamma
N\Delta}}{g_{\rm \pi NN}}\frac{g^3_{\rm \pi NN}}{4M^2_{\pi}}\int
d^4x_1 d^4x_2 d^4x_3\int
d^4z\,\frac{\partial}{\partial z_{\varphi}}\delta^{(4)}(z-x_1)\nonumber\\
\hspace{-0.5in}&&\times\,\Big\{{\rm T}([\bar{p^c}(x_1)\gamma^5 n(z) +
\bar{p^c}(z)\gamma^5
n(x_1)]\,D^{\dagger}_{\mu}(x_2)
F^{\alpha\beta}(x_3))\nonumber\\
\hspace{-0.5in}&&\times\,\frac{1}{i}{\rm tr}\{S_{\alpha\omega}(x_3 - z)\,{\Theta^{\omega}}_{\varphi}\, S^c_{\rm F}(x_1 - x_2) \gamma^{\mu}
S_{\rm F}(x_2 - x_3) \gamma_{\beta}\gamma^5\}\nonumber\\
\hspace{-0.5in}&&-{\rm T}([\bar{p^c}(x_1)\gamma_{\nu}\gamma^5 n(z) +
\bar{p^c}(z)\gamma_{\nu}\gamma^5
n(x_1)]\,D^{\dagger}_{\mu}(x_2)
F^{\alpha\beta}(x_3))\nonumber\\
\hspace{-0.5in}&&\times\,\frac{1}{i}{\rm tr}\{S_{\alpha\omega}(x_3 - z) \, {\Theta^{\omega}}_{\varphi}\,\gamma^{\nu} S^c_{\rm F}(x_1 - x_2) \gamma^{\mu} S_{\rm F}(x_2 - x_3) \gamma_{\beta}\gamma^5\}\Big\}.
\end{eqnarray}
Thus, the effective Lagrangian ${\cal L}_{\rm np \to \Delta N \to
D\gamma}(x)$ reads
\begin{eqnarray}\label{label3.10}
\hspace{-0.5in}&&\int d^4x\,{\cal L}_{\rm np \to \Delta N \to
D\gamma}(x) = \frac{2}{3}\frac{ieg_{\rm V}}{M^2_{\rm N}}\frac{g_{\rm
\pi N\Delta}}{g_{\rm \pi NN}}\frac{g_{\rm \gamma N\Delta}}{g_{\rm \pi
NN}}\frac{g^3_{\rm \pi NN}}{4M^2_{\pi}}\int d^4x_1 d^4x_2 d^4x_3\int d^4z\,\frac{\partial}{\partial z_{\varphi}}\delta^{(4)}(z-x_1)\nonumber\\
\hspace{-0.5in}&&\times\,\Big\{{\rm T}([\bar{p^c}(x_1)\gamma^5 n(z) +
\bar{p^c}(z)\gamma^5 n(x_1)]\, D^{\dagger}_{\mu}(x_2) F^{\alpha\beta}(x_3))\nonumber\\
\hspace{-0.7in}&&\times\,\frac{1}{i}{\rm tr}\{S_{\alpha\omega}(x_3-z)\, {\Theta^{\omega}}_{\varphi}\, S^c_{\rm F}(x_1-x_2) \gamma^{\mu} S_{\rm F}(x_2 - x_3) \gamma_{\beta}\gamma^5\}\nonumber\\
\hspace{-0.5in}&&-{\rm T}([\bar{p^c}(x_1)\gamma_{\nu}\gamma^5 n(z) +
\bar{p^c}(z)\gamma_{\nu}\gamma^5
n(x_1)]\, D^{\dagger}_{\mu}(x_2) F^{\alpha\beta}(x_3))\nonumber\\
\hspace{-0.5in}&&\times\,\frac{1}{i}{\rm
tr}\{S_{\alpha\omega}(x_3-z)\,
{\Theta^{\omega}}_{\varphi}\,\gamma^{\nu} S^c_{\rm F}(x_1-x_2)
\gamma^{\mu} S_{\rm F}(x_2 - x_3) \gamma_{\beta}\gamma^5\}\Big\}.
\end{eqnarray}
In the momentum representation of the baryon Green functions the
effective Lagrangian Eq.(\ref{label3.10}) reads
\begin{eqnarray}\label{label3.11}
\hspace{-0.5in}&&\int d^4x\,{\cal L}_{\rm np \to \Delta N \to
D\gamma}(x) = \frac{2}{3}\frac{ieg_{\rm V}}{M^2_{\rm N}}\frac{g_{\rm
\pi N\Delta}}{g_{\rm \pi NN}}\frac{g_{\rm \gamma N\Delta}}{g_{\rm \pi
NN}}\frac{g^3_{\rm \pi NN}}{4M^2_{\pi}}\int d^4x_1\int
d^4z\,\frac{\partial}{\partial z_{\varphi}}\delta^{(4)}(z-x_1)
\nonumber\\
\hspace{-0.5in}&&\times\,\int\frac{d^4x_2d^4k_2}{(2\pi)^4}
\frac{d^4x_3d^4k_3}{(2\pi)^4}\,e^{\textstyle -ik_2\cdot (x_2 - x_1)}\,e^{\textstyle -ik_3\cdot
(x_3 - z)}\nonumber\\
\hspace{-0.5in}&&\times\Big\{{\rm T}([\bar{p^c}(x_1)\gamma^5 n(z) + \bar{p^c}(z)\gamma^5 n(x_1)]\, D^{\dagger}_{\mu}(x_2) F_{\alpha\beta}(x_3))\nonumber\\
\hspace{-0.5in}&&\times\,\int\frac{d^4k_1}{\pi^2i}\,e^{\textstyle
ik_1\cdot (x_1 - z)}\,{\rm tr}\{S^{\alpha\omega}(k_1 + k_3)\,
\Theta_{\omega\varphi}\,\frac{1}{M_{\rm N} - \hat{k}_1 + \hat{k}_2} \gamma^{\mu}
\frac{1}{M_{\rm N} - \hat{k}_1}
\gamma^{\beta}\gamma^5\}\nonumber\\
\hspace{-0.5in}&&-{\rm T}([\bar{p^c}(x_1)\gamma_{\nu}\gamma^5 n(z) +
\bar{p^c}(z)\gamma_{\nu}\gamma^5
n(x_1)] \, D^{\dagger}_{\mu}(x_2) F^{\alpha\beta}(x_3))\nonumber\\
\hspace{-0.5in}&&\times\,\int\frac{d^4k_1}{\pi^2i}\,e^{\textstyle
ik_1\cdot (x_1 - z)}\,{\rm
tr}\{S^{\alpha\omega}(k_1 + k_3)\,
\Theta_{\omega\varphi}\,\frac{1}{M_{\rm N} - \hat{k}_1 + \hat{k}_2}
\gamma^{\mu} \frac{1}{M_{\rm N} -
\hat{k}_1}\gamma^{\beta}\gamma^5\}\Big\}.
\end{eqnarray}
The effective Lagrangian Eq.(\ref{label3.11}) defines the contribution
of the $\Delta$ resonance to the amplitude of the neutron--proton
radiative capture.

The amplitude of the neutron--proton radiative capture caused by the
contribution of the $\Delta(1232)$ resonance exchange we define by a usual
way [5]:
\begin{eqnarray}\label{label3.12}
\hspace{-0.5in}&&\int d^4x\,<{\rm D}(k_{\rm D}) \gamma(k)|{\cal
L}_{\rm np \to \Delta N \to D\gamma}(x)|n(p_1)p(p_2)>=\nonumber\\
\hspace{-0.5in}&&= (2\pi)^4\delta^{(4)}(k_{\rm D} + k - p_1 - p_2)\,\frac{{\cal M}({\rm n + p \to \Delta N \to D + \gamma})}{\displaystyle \sqrt{2 E_1 V\,2 E_2 V\,2 E_{\rm D} V\,2 \omega V}},
\end{eqnarray}
where $E_i\,(i=1,2,{\rm D})$ and $\omega$ are the energies of the
neutron, the proton, the deuteron and the photon, and $V$ is the
normalization volume. For the computation of the amplitude ${\cal
M}({\rm n + p \to \Delta N \to D + \gamma})$ we should take the
r.h.s. of Eq.(\ref{label3.11}) between the wave functions of the
initial $|n(p_1)p(p_2)>$ and the final $<{\rm D}(k_{\rm D})\gamma(k)|$
states. This gives
\begin{eqnarray}\label{label3.13}
\hspace{-0.5in}&& (2\pi)^4\delta^{(4)}(k_{\rm D} + k - p_1 -
p_2)\,\frac{{\cal M}({\rm n + p \to \Delta N \to D +
\gamma})}{\displaystyle \sqrt{2 E_1 V\,2 E_2 V\,2 E_{\rm D} V\,2
\omega V}}=\nonumber\\
\hspace{-0.5in}&& = \frac{2}{3}\frac{ieg_{\rm V}}{M^2_{\rm
N}}\frac{g_{\rm \pi N\Delta}}{g_{\rm \pi NN}}\frac{g_{\rm \gamma
N\Delta}}{g_{\rm \pi NN}}\frac{g^3_{\rm \pi NN}}{4M^2_{\pi}}\int
d^4x_1 \int d^4z\,\frac{\partial}{\partial
z_{\varphi}}\delta^{(4)}(z-x_1)\nonumber\\
\hspace{-0.5in}&&\times\, \int\frac{d^4x_2d^4k_2}{(2\pi)^4}
\frac{d^4x_3d^4k_3}{(2\pi)^4}\,e^{\textstyle - ik_2\cdot (x_2 - x_1)}\,e^{\textstyle -ik_3\cdot
(x_3 - z)} \nonumber\\
\hspace{-0.5in}&&\times \Big\{<{\rm D}(k_{\rm D})\gamma(k)|{\rm
T}([\bar{p^c}(x_1)\gamma^5 n(z) + \bar{p^c}(z)\gamma^5 n(x_1)]\,
D^{\dagger}_{\mu}(x_2) F_{\alpha\beta}(x_3))|n(p_1)p(p_2)>\nonumber\\
\hspace{-0.5in}&&\times\,\int\frac{d^4k_1}{\pi^2i}\,e^{\textstyle
ik_1\cdot (x_1 - z)}\,{\rm tr}\{S^{\alpha\omega}(k_1 + k_3)\,
\Theta_{\omega\varphi}\,\frac{1}{M_{\rm N} - \hat{k}_1 + \hat{k}_2}
\gamma^{\mu} \frac{1}{M_{\rm N} - \hat{k}_1}
\gamma^{\beta}\gamma^5\}\nonumber\\
\hspace{-0.5in}&&-
<{\rm D}(k_{\rm D})\gamma(k)|{\rm T}([\bar{p^c}(x_1)\gamma_{\nu}
\gamma^5 n(z) + \bar{p^c}(z)\gamma_{\nu}\gamma^5 n(x_1)]\, 
D^{\dagger}_{\mu}(x_2) F_{\alpha\beta}(x_3))|n(p_1)p(p_2)>\nonumber\\
\hspace{-0.5in}&&\times\,\int\frac{d^4k_1}{\pi^2i}\,e^{\textstyle
ik_1\cdot (x_1 - z)}\,{\rm
tr}\{S^{\alpha\omega}(k_1 + k_3)\,
\Theta_{\omega\varphi}\,\frac{1}{M_{\rm N} - \hat{k}_1 + \hat{k}_2} \gamma^{\mu}
\frac{1}{M_{\rm N} - \hat{k}_1}
\gamma^{\beta}\gamma^5\}\Big\}.
\end{eqnarray}
The matrix elements between the initial and the final states are given
by [5]:
\begin{eqnarray}\label{label3.14}
\hspace{-0.5in}&&<{\rm D}(k_{\rm D})\gamma(k)|{\rm
T}([\bar{p^c}x_1)\gamma^5 n(z) +
\bar{p^c}(z)\gamma^5
n(x_1)]\, D^{\dagger}_{\mu}(x_2)
F_{\alpha\beta}(x_3))|n(p_1)p(p_2)>=\nonumber\\
\hspace{-0.5in}&&= [\bar{u^c}(p_2)\gamma^5
u(p_1)]\times\,i\,(k_{\alpha} e^*_{\beta}(k) - k_{\beta}
e^*_{\alpha}(k)\times\,e^*_{\mu}(k_{\rm
D})\nonumber\\
\hspace{-0.5in}&&\times\,\frac{e^{\textstyle i k_{\rm D}\cdot
x_2}\,e^{\textstyle i k\cdot x_3}}{\displaystyle \sqrt{2 E_1 V\,2 E_2
V\,2 E_{\rm D} V\,2 \omega V}}\,\Bigg(e^{\textstyle - i p_1\cdot x_1 -
i p_2\cdot z}+ e^{\textstyle - i p_2\cdot x_1 -
i p_1\cdot z}\Bigg)\,,\nonumber\\
\hspace{-0.5in}&&<{\rm D}(k_{\rm D})\gamma(k)|{\rm
T}([\bar{p^c}(x_1)\gamma_{\nu}\gamma^5 n(z) +
\bar{p^c}(z)\gamma_{\nu}\gamma^5 n(x_1)]\, D^{\dagger}_{\mu}(x_2)
F_{\alpha\beta}(x_3))|n(p_1)p(p_2)>=\nonumber\\
\hspace{-0.5in}&&=[\bar{u^c}(p_2)\gamma_{\nu}\gamma^5
u(p_1)]\times\,i\,(k_{\alpha} e^*_{\beta}(k) - k_{\beta}
e^*_{\alpha}(k))\times\,e^*_{\mu}(k_{\rm
D})\nonumber\\
\hspace{-0.5in}&&\times\,\frac{e^{\textstyle i k_{\rm D}\cdot
x_2}\,e^{\textstyle i k\cdot x_3}}{\displaystyle \sqrt{2 E_1 V\,2
E_2 V\,2 E_{\rm D} V\,2 \omega V}}\,\Bigg(e^{\textstyle - i p_1\cdot
x_1 - i p_2\cdot z}+ e^{\textstyle - i p_2\cdot x_1 - i p_1\cdot
z}\Bigg).
\end{eqnarray}
Substituting Eq.(\ref{label3.14}) in Eq.(\ref{label3.13}) and
integrating over $z$ we obtain
\begin{eqnarray}\label{label3.15}
\hspace{-0.5in}&& (2\pi)^4\delta^{(4)}(k_{\rm D} + k - p_1 -
p_2)\,{\cal M}({\rm n + p \to \Delta N \to D + \gamma}) =\nonumber\\
\hspace{-0.5in}&& = -\frac{ie}{2M^2_{\rm N}}\frac{g_{\rm
V}}{6\pi^2}\frac{g_{\rm \pi N\Delta}}{g_{\rm \pi NN}}\frac{g_{\rm
\gamma N\Delta}}{g_{\rm \pi NN}}\frac{g^3_{\rm \pi
NN}}{4M^2_{\pi}}[\bar{u^c}(p_2)\gamma^5 u(p_1)]\,(k_{\alpha}
e^*_{\beta}(k) - k_{\beta} e^*_{\alpha}(k))\,e^*_{\mu}(k_{\rm
D})\nonumber\\
\hspace{-0.5in}&&\times\,\int d^4x_1 \int\frac{d^4x_2d^4k_2}{(2\pi)^4}
\frac{d^4x_3d^4k_3}{(2\pi)^4}\,e^{\textstyle i(k_2 + k_3 - p_1 -
p_2)\cdot x_1}\,e^{\textstyle i (k_{\rm D} - k_2)\cdot
x_2}\,e^{\textstyle i (k - k_3)\cdot x_3} \nonumber\\
\hspace{-0.5in}&&\times\,\int\frac{d^4k_1}{\pi^2i}\,{\rm tr}\{(p_1 +
p_2 + k_1 - k_3)^{\varphi}S^{\alpha\omega}(k_1 + k_3)\,
\Theta_{\omega\varphi}\,\frac{1}{M_{\rm N} - \hat{k}_1 + \hat{k}_2}
\gamma^{\mu} \frac{1}{M_{\rm N} - \hat{k}_1}
\gamma_{\beta}\gamma^5\}\nonumber\\
\hspace{-0.5in}&&+\frac{ie}{2M^2_{\rm N}}\frac{g_{\rm
V}}{6\pi^2}\frac{g_{\rm \pi N\Delta}}{g_{\rm \pi NN}}\frac{g_{\rm
\gamma N\Delta}}{g_{\rm \pi NN}}\frac{g^3_{\rm \pi
NN}}{4M^2_{\pi}}[\bar{u^c}(p_2)\gamma_{\nu}\gamma^5
u(p_1)]\,(k_{\alpha} e^*_{\beta}(k) - k_{\beta}
e^*_{\alpha}(k))\,e^*_{\mu}(k_{\rm D})\nonumber\\
\hspace{-0.5in}&&\times\,\int d^4x_1 \int\frac{d^4x_2d^4k_2}{(2\pi)^4}
\frac{d^4x_3d^4k_3}{(2\pi)^4}\,e^{\textstyle i(k_2 + k_3 - p_1 -
p_2)\cdot x_1}\,e^{\textstyle i (k_{\rm D} - k_2)\cdot
x_2}\,e^{\textstyle i (k - k_3)\cdot x_3} \nonumber\\
\hspace{-0.5in}&&\times\,\int\frac{d^4k_1}{\pi^2i}\,{\rm tr}\{(p_1 +
p_2 + k_1 - k_3)^{\varphi}S^{\alpha\omega}(k_1 + k_3)\,
\Theta_{\omega\varphi}\,\gamma^{\nu}\frac{1}{M_{\rm N} - \hat{k}_1 +
\hat{k}_2} \gamma^{\mu} \frac{1}{M_{\rm N} - \hat{k}_1}
\gamma_{\beta}\gamma^5\}.\nonumber\\
\hspace{-0.5in}&&
\end{eqnarray}
Integrating over $x_1$, $x_2$, $x_3$, $k_2$ and $k_3$ we obtain in the
r.h.s. of Eq.(\ref{label3.15}) the $\delta$--function describing the
4--momentum conservation. The, the amplitude ${\cal M}({\rm n + p \to
\Delta N \to D + \gamma})$ becomes equal
\begin{eqnarray}\label{label3.16}
\hspace{-0.5in}&& {\cal M}({\rm n + p \to \Delta N \to D + \gamma})
=\nonumber\\
\hspace{-0.5in}&& = -\frac{ie}{2M^2_{\rm N}}\frac{g_{\rm
V}}{6\pi^2}\frac{g_{\rm \pi N\Delta}}{g_{\rm \pi NN}}\frac{g_{\rm
\gamma N\Delta}}{g_{\rm \pi NN}}\frac{g^3_{\rm \pi
NN}}{4M^2_{\pi}}[\bar{u^c}(p_2)\gamma^5 u(p_1)]\,(k_{\alpha}
e^*_{\beta}(k) - k_{\beta} e^*_{\alpha}(k))\,e^*_{\mu}(k_{\rm
D})\nonumber\\
\hspace{-0.5in}&&\times\,\int\frac{d^4k_1}{\pi^2i}\,{\rm tr}\{(k_1 +
k_{\rm D})^{\varphi}S^{\alpha\omega}(k_1 + k)\,
\Theta_{\omega\varphi}\,\frac{1}{M_{\rm N} - \hat{k}_1+ \hat{k}_{\rm
D}} \gamma^{\mu} \frac{1}{M_{\rm N} - \hat{k}_1 }
\gamma_{\beta}\gamma^5\}\nonumber\\
\hspace{-0.5in}&&+\frac{e}{2M^2_{\rm N}}\frac{g_{\rm
V}}{6\pi^2}\frac{g_{\rm \pi N\Delta}}{g_{\rm \pi NN}}\frac{g_{\rm
\gamma N\Delta}}{g_{\rm \pi NN}}\frac{g^3_{\rm \pi
NN}}{4M^2_{\pi}}[\bar{u^c}(p_2)\gamma_{\nu}\gamma^5
u(p_1)]\,(k_{\alpha} e^*_{\beta}(k) - k_{\beta}
e^*_{\alpha}(k))\,e^*_{\mu}(k_{\rm D})\nonumber\\
\hspace{-0.5in}&&\times\,\int\frac{d^4k_1}{\pi^2i}\,{\rm tr}\{(k_1 +
k_{\rm D})^{\varphi}S^{\alpha\omega}(k_1 + k)\,
\Theta_{\omega\varphi}\,\gamma^{\nu}\frac{1}{M_{\rm N} - \hat{k}_1 +
\hat{k}_{\rm D}} \gamma^{\mu} \frac{1}{M_{\rm N} - \hat{k}_1}
\gamma_{\beta}\gamma^5\},
\end{eqnarray}
For the subsequent calculation it is convenient to introduce the
structure functions
\begin{eqnarray}\label{label3.17}
\hspace{-0.5in}&& {\cal M}({\rm n + p \to \Delta N \to D + \gamma})
=\nonumber\\
\hspace{-0.5in}&& = -\frac{ie}{2M^2_{\rm N}}\frac{g_{\rm
V}}{6\pi^2}\frac{g_{\rm \pi N\Delta}}{g_{\rm \pi NN}}\frac{g_{\rm
\gamma N\Delta}}{g_{\rm \pi NN}}\frac{g^3_{\rm \pi
NN}}{4M^2_{\pi}}[\bar{u^c}(p_2)\gamma^5 u(p_1)]\,(k_{\alpha}
e^*_{\beta}(k) - k_{\beta} e^*_{\alpha}(k))\,e^*_{\mu}(k_{\rm
D})\nonumber\\
\hspace{-0.5in}&&\times\,{\cal J}^{\mu\beta\alpha}_5(k_{\rm
D},k)\nonumber\\
\hspace{-0.5in}&&+\frac{ie}{2M^2_{\rm N}}\frac{g_{\rm
V}}{6\pi^2}\frac{g_{\rm \pi N\Delta}}{g_{\rm \pi NN}}\frac{g_{\rm
\gamma N\Delta}}{g_{\rm \pi NN}}\frac{g^3_{\rm \pi
NN}}{4M^2_{\pi}}[\bar{u^c}(p_2)\gamma_{\nu}\gamma^5
u(p_1)]\,(k_{\alpha} e^*_{\beta}(k) - k_{\beta}
e^*_{\alpha}(k))\,e^*_{\mu}(k_{\rm D})\nonumber\\
\hspace{-0.5in}&&\times\,{\cal J}^{\nu\mu\beta\alpha}_5(k_{\rm D},k),
\end{eqnarray}
where the structure functions ${\cal J}^{\mu\beta\alpha}_5(k_{\rm
D},k)$ and ${\cal J}^{\nu\mu\beta\alpha}_5(k_{\rm D},k)$ are defined
by the momentum integrals
\begin{eqnarray}\label{label3.18}
\hspace{-0.5in}&& {\cal J}^{\mu\beta\alpha}_5(k_{\rm D},k) =\nonumber\\
\hspace{-0.5in}&&=\int\frac{d^4k_1}{\pi^2i}\,{\rm tr}\{(k_1 + k_{\rm
D})^{\varphi}S^{\alpha\omega}(k_1 + k)\,
\Theta_{\omega\varphi}\,\frac{1}{M_{\rm N} - \hat{k}_1+ \hat{k}_{\rm
D}} \gamma^{\mu} \frac{1}{M_{\rm N} - \hat{k}_1 }
\gamma^{\beta}\gamma^5\},\nonumber\\
\hspace{-0.5in}&&{\cal J}^{\nu\mu\beta\alpha}_5(k_{\rm D},k) =\nonumber\\
\hspace{-0.5in}&&=\int\frac{d^4k_1}{\pi^2i}\,{\rm tr}\{(k_1 + k_{\rm
D})^{\varphi}S^{\alpha\omega}(k_1 + k)\,
\Theta_{\omega\varphi}\,\gamma^{\nu}\frac{1}{M_{\rm N} - \hat{k}_1 +
\hat{k}_{\rm D}} \gamma^{\mu} \frac{1}{M_{\rm N} - \hat{k}_1}
\gamma^{\beta}\gamma^5\}.
\end{eqnarray}
Now the problem of the calculation of the contribution of the
$\Delta(1232)$ resonance to the amplitude of the neutron--proton
radiative capture is reduced to the problem of the evaluation of the
structure functions. In the leading order in large $N_C$ expansion [1]
we obtain
\begin{eqnarray}\label{label3.19}
{\cal J}^{\mu\beta\alpha}_5(k_{\rm D},k)&=& \frac{4}{3}\,\Bigg(Z
-\frac{1}{2}\Bigg)\,i\,M_{\rm N}\,
\varepsilon^{\mu\beta\alpha\lambda}\,k_{\rm D\lambda},\nonumber\\
{\cal J}^{\nu\mu\beta\alpha}_5(k_{\rm D},k)&=& \frac{2}{3}\,\Bigg(Z
-\frac{1}{2}\Bigg)\,i\,M^2_{\rm N}\, \varepsilon^{\mu\beta\alpha\nu}.
\end{eqnarray}
We have neglected the mass difference between the masses of the
$\Delta(1232)$ resonance and the nucleon.  The amplitude of the
neutron--proton radiative capture caused by the $\Delta(1232)$
resonance contribution is equal to
\begin{eqnarray}\label{label3.20}
\hspace{-0.5in}&&{\cal M}_{\Delta}({\rm n + p \to D + \gamma}) =
\frac{e}{2M_{\rm N}}\frac{g_{\rm V}}{4\pi^2}\Bigg[\Bigg(\frac{1}{2} -
Z\Bigg)\,\frac{8}{9}\,\frac{g_{\rm \pi N\Delta}}{g_{\rm \pi
NN}}\,\frac{g_{\rm \gamma N\Delta}}{g_{\rm \pi NN}}\,\frac{g^3_{\rm
\pi NN}}{4M^2_{\pi}}\Bigg]\, \nonumber\\
\hspace{-0.5in}&&\times\, \varepsilon^{\alpha\beta\mu\nu} k_{\alpha}
e^*_{\beta}(k) e^*_{\mu}(k_{\rm D}) [\bar{u^c}(p_2)(2k_{\rm D\nu} -
M_{\rm N}\gamma_{\nu})\gamma^5 u(p_1)]=\nonumber\\
\hspace{-0.5in}&&=e \,\frac{5 g_{\rm V}}{8\pi^2}\,\Bigg[\Bigg(\frac{1}{2}
- Z\Bigg)\,\frac{8}{9}\,\frac{g_{\rm \pi N\Delta}}{g_{\rm \pi
NN}}\,\frac{g_{\rm \gamma N\Delta}}{g_{\rm \pi NN}}\,\frac{g^3_{\rm \pi
NN}}{4M^2_{\pi}}\Bigg] (\vec{k}\times \vec{e}^{\,*}(\vec{k}\,))\cdot
\vec{e}^{\,*}(\vec{k}_{\rm D}) \,[\bar{u^c}(p_2)\gamma^5 u(p_1)].
\end{eqnarray}
The total amplitude of the neutron--proton radiative capture for
thermal neutrons reads
\begin{eqnarray}\label{label3.21}
\hspace{-0.7in}&&{\cal M}({\rm n + p \to D + \gamma}) =  e\, (\mu_{\rm
p} - \mu_{\rm n}) \frac{5 g_{\rm V}}{8\pi^2} G_{\rm \pi NN}
(\vec{k}\times \vec{e}^{\,*}(\vec{k}\,))\cdot
\vec{e}^{\,*}(\vec{k}_{\rm D}) \,[\bar{u^c}(p_2)\gamma^5 u(p_1)]\nonumber\\
\hspace{-0.7in}&&\times\,\Bigg[1 + \frac{g^2_{\rm \pi NN}}{\mu_{\rm p}
- \mu_{\rm
n}}\,\frac{M^2_{\pi}}{8\pi^2}\,\frac{\alpha_{\rho}}{\pi}\Bigg(J_{\rm
\pi a_1 N} + \frac{3}{2g_{\rm A}}\,J_{\rm
\pi VN}\Bigg) + \frac{1 - 2Z}{\mu_{\rm p} - \mu_{\rm
n}}\,\frac{1}{G_{\rm \pi NN}}\,\frac{4}{9}\,\frac{g_{\rm \pi
N\Delta}}{g_{\rm \pi NN}}\,\frac{g_{\rm \gamma N\Delta}}{g_{\rm \pi
NN}}\,\frac{g^3_{\rm \pi NN}}{4M^2_{\pi}}\Bigg].
\end{eqnarray}
The total cross section for the neutron--proton radiative capture is
then defined by
\begin{eqnarray}\label{label3.22}
\hspace{-0.7in}&&\sigma^{\rm n p}(k)=
\frac{1}{v}\,(\mu_{\rm p}-\mu_{\rm
n})^2\,\frac{25}{64}\,\frac{\alpha}{\pi^2}\,\,Q_{\rm D}\,G^2_{\rm \pi
NN}\,M_{\rm N}\,\varepsilon^3_{\rm D}\,\nonumber\\
\hspace{-0.7in}&&\times\,\Bigg[1 + \frac{g^2_{\rm \pi NN}}{\mu_{\rm p}
- \mu_{\rm
n}}\,\frac{M^2_{\pi}}{8\pi^2}\,\frac{\alpha_{\rho}}{\pi}\Bigg(J_{\rm
\pi a_1 N} + \frac{3}{2 g_{\rm A}}\,J_{\rm \pi VN}\Bigg) + \frac{1 -
2Z}{\mu_{\rm p} - \mu_{\rm n}}\,\frac{1}{G_{\rm \pi
NN}}\,\frac{4}{9}\,\frac{g_{\rm \pi N\Delta}}{g_{\rm \pi
NN}}\,\frac{g_{\rm \gamma N\Delta}}{g_{\rm \pi NN}}\,\frac{g^3_{\rm
\pi NN}}{4M^2_{\pi}}\Bigg]^2.
\end{eqnarray}
The numerical value of the cross section amounts to
\begin{eqnarray}\label{label3.23}
\sigma^{\rm n p}(k)= 287.2\,(1 + 0.64\,(1-2Z))^2\,{\rm m\,b}.
\end{eqnarray}
Thus, the discrepancy of the theoretical cross section and the
experimental value $\sigma^{\rm n p}_{\exp}\,=\,(334.2\pm 0.5)\,{\rm
mb}$ can by described by the contribution of the $\Delta(1232)$
resonance. In order to fit the experimental value of the cross section
we should take $Z$ equal to $Z=0.438$. This agrees with the
experimental bound $|Z|\le 1/2$ [23].

\section{Photo--magnetic disintegration of the deuteron}
\setcounter{equation}{0}

\hspace{0.2in} The amplitude of the photo--magnetic disintegration of
the deuteron $\gamma$ + D $\to$ n + p evaluated near threshold is
related to the amplitude of the neutron--proton radiative capture n +
p $\to$ D + $\gamma$ and reads
\begin{eqnarray}\label{label4.1}
\hspace{-0.5in}&&{\cal M}({\rm \gamma + D \to n + p}) = e\,(\mu_{\rm p} - \mu_{\rm
n})\,\frac{5 g_{\rm V}}{8\pi^2}\,G_{\rm \pi NN}\,(\vec{q}\times
\vec{e}(\vec{q}\,))\cdot \vec{e}(\vec{k}_{\rm D})
\,[\bar{u}(p_2)\gamma^5 u^c(p_1)]\,\nonumber\\ 
\hspace{-0.5in}&&\times\,\Bigg[1 + \frac{g^2_{\rm \pi NN}}{\mu_{\rm p}
- \mu_{\rm
n}}\,\frac{M^2_{\pi}}{8\pi^2}\,\frac{\alpha_{\rho}}{\pi}\Bigg(J_{\rm
\pi a_1 N} + \frac{3}{2 g_{\rm A}}\,J_{\rm \pi VN}\Bigg) +
\frac{1 - 2Z}{\mu_{\rm p} - \mu_{\rm n}}\,\frac{1}{G_{\rm \pi
NN}}\,\frac{4}{9}\,\frac{g_{\rm \pi N\Delta}}{g_{\rm \pi
NN}}\,\frac{g_{\rm \gamma N\Delta}}{g_{\rm \pi NN}}\,\frac{g^3_{\rm
\pi NN}}{4M^2_{\pi}}\Bigg].
\end{eqnarray}
The cross section defined by the amplitude Eq.(\ref{label4.1}) is then
given by
\begin{eqnarray}\label{label4.2}
\sigma^{\rm \gamma D}(\omega) = \sigma_0
\,\Bigg(\frac{\omega}{\varepsilon_{\rm D}}\Bigg)\,
k r_{\rm D},
\end{eqnarray}
where $k=\sqrt{M_{\rm N}(\omega - \varepsilon_{\rm D})}$ is the
relative momentum of the np system, $\omega$ is the energy of the
photon and $r_{\rm D} = 1/\sqrt{\varepsilon_{\rm D}M_{\rm N}} =
4.315\,{\rm fm}$ is the radius of the deuteron, and $\sigma_0$ is
equal to
\begin{eqnarray}\label{label4.3}
\hspace{-0.3in}&&\sigma_0 = (\mu_{\rm p}-\mu_{\rm n})^2 \frac{25\alpha
Q_{\rm D}}{192\pi^2} G^2_{\rm \pi NN}\,\varepsilon^{3/2}_{\rm
D}M^{5/2}_{\rm N}\,\nonumber\\
\hspace{-0.3in}&&\times\,\Bigg[1 + \frac{g^2_{\rm \pi NN}}{\mu_{\rm p}
- \mu_{\rm
n}}\,\frac{M^2_{\pi}}{8\pi^2}\,\frac{\alpha_{\rho}}{\pi}\Bigg(J_{\rm
\pi a_1 N} + \frac{3}{2 g_{\rm A}}\,J_{\rm \pi VN}\Bigg) +
\frac{1 - 2Z}{\mu_{\rm p} - \mu_{\rm n}}\,\frac{1}{G_{\rm \pi
NN}}\,\frac{4}{9}\,\frac{g_{\rm \pi N\Delta}}{g_{\rm \pi
NN}}\,\frac{g_{\rm \gamma N\Delta}}{g_{\rm \pi NN}}\,\frac{g^3_{\rm
\pi NN}}{4M^2_{\pi}}\Bigg]^2=\nonumber\\
\hspace{-0.6in}&&\quad\, = 7.10\,{\rm m\,b}.
\end{eqnarray}
The cross section $\sigma^{\rm \gamma D}(\omega)$ calculated in the
PMA near threshold has the same form as Eq.\,(\ref{label4.2}) but with
$\sigma_0$ amounting to [17]
\begin{eqnarray}\label{label4.4}
\hspace{-0.5in} \sigma_0 = \frac{2\pi\alpha}{3 M^2_{\rm N}}\,(\mu_{\rm
p}-\mu_{\rm n})^2
\Big(1 - a_{\rm np}\sqrt
{\varepsilon_{\rm D} M_{\rm N}}\Big)^2 = 6.31\,{\rm m b}.
\end{eqnarray}
It is seen that numerical values of $\sigma_0$ defined by
Eq.\,(\ref{label4.3}) and Eq.\,(\ref{label4.4}) are in good
agreement. We should emphasize that the cross section
Eq.(\ref{label4.4}) does not contain corrections mentioned by Riska
and Brown [11] (see also [17]) increasing its value.

In order to obtain the cross section for the process $\gamma$ + D
$\to$ n + p far from threshold we take into account the np interaction in
the final state. This can be carried out by summing up an infinite
series of one--nucleon loop diagrams. In this case the amplitude of
$\gamma$ + D $\to$ n + p reads
\begin{eqnarray}\label{label4.5}
\hspace{-0.5in} &&{\cal M}({\rm \gamma + D \to n + p}) = {\cal A}_{\rm
th}\, [\bar{u}(p_2)\gamma^5 u^c(p_1)]\nonumber\\
\hspace{-0.5in} &&\times \frac{1}{\displaystyle 1 + \frac{G_{\rm \pi
NN}}{16\pi^2}\int \frac{d^4p}{\pi^2i}\,{\rm tr}\Bigg\{\gamma^5
\frac{1}{M_{\rm N} - \hat{p} - \hat{P} - \hat{Q}}\gamma^5
\frac{1}{M_{\rm N} - \hat{p} - \hat{Q}}\Bigg\}},
\end{eqnarray}
where ${\cal A}_{\rm th}$ is the amplitude calculated near threshold,
$P = p_1 + p_2 = (2\sqrt{k^2 + M^2_{\rm N}}, \vec{0}\,)$ is the
4--momentum of the np system in the center of mass frame. Then, $Q
=a\,P + b\,K = a\,(p_1 + p_2) + b\,(p_1 - p_2)$ is an arbitrary shift
of virtual momentum with arbitrary parameters $a$ and $b$, and  in the
center of mass frame $K = p_1 - p_2 = (0,2\,\vec{k}\,)$. The explicit
dependence of the momentum integral on $Q$ can be evaluated by means
of the Gertsein--Jackiw procedure [28] and is given by [1--6]:
\begin{eqnarray}\label{label4.6}
\hspace{-0.5in} &&\int \frac{d^4p}{\pi^2i}\,{\rm tr}\Bigg\{\gamma^5
\frac{1}{M_{\rm N} - \hat{p} - \hat{P} - \hat{Q}}\gamma^5
\frac{1}{M_{\rm N} - \hat{p} - \hat{Q}}\Bigg\} =\nonumber\\
\hspace{-0.5in} &&=\int \frac{d^4p}{\pi^2i}\,{\rm tr}\Bigg\{\gamma^5
\frac{1}{M_{\rm N} - \hat{p} - \hat{P}}\gamma^5 \frac{1}{M_{\rm N} -
\hat{p}}\Bigg\} - 2\, a\,(a + 1)\,P^2 - 2\,b^2\,K^2.
\end{eqnarray}
For the evaluation of the momentum integral over $p$ we would keep
only the leading order contributions in the large $N_C$ expansion
[1]. This yields
\begin{eqnarray}\label{label4.7}
\hspace{-0.5in} &&\int \frac{d^4p}{\pi^2i}\,{\rm tr}\Bigg\{\gamma^5
\frac{1}{M_{\rm N} - \hat{p} - \hat{P} - \hat{Q}}\gamma^5
\frac{1}{M_{\rm N} - \hat{p} - \hat{Q}}\Bigg\} =\nonumber\\
\hspace{-0.5in} &&=- 8\, a\,(a + 1)\,M^2_{\rm N} + 8\,(b^2 - a\,(a +
1))\,k^2 - i\,8\pi\,M_{\rm N}\,k.
\end{eqnarray}
The amplitude Eq.(\ref{label4.5}) we obtain in the form
\begin{eqnarray}\label{label4.8}
{\cal M}({\rm \gamma + D \to n + p}) = {\cal A}_{\rm th}\,
[\bar{u}(p_2)\gamma^5 u^c(p_1)]\, \frac{Z}{\displaystyle 1 -
\frac{1}{2} r_{\rm np} a_{\rm np} k^2 + i a_{\rm np} k}.
\end{eqnarray}
Here we have denoted 
\begin{eqnarray}\label{label4.9}
a_{\rm np} &=& - \frac{G_{\rm \pi NN}M_{\rm N}}{2\pi}\,Z\quad,\quad
r_{\rm np} = (b^2 - a\,(a + 1))\,\frac{2}{\pi}\,\frac{1}{M_{\rm N}},
\nonumber\\ \frac{1}{Z}&=& 1 - \frac{a(a+1)}{2\pi^2}\,G_{\rm \pi
NN}\,M^2_{\rm N},
\end{eqnarray}
where $r_{\rm np} = 2.75\pm 0.05\,{\rm fm}$ is the effective range of
 low--energy elastic np scattering.

  Renormalizing the wave functions of nucleons $\sqrt{Z}\,u(p_1) \to
u(p_1)$ and $\sqrt{Z}\,u(p_2) \to u(p_2)$ we arrive at the amplitude
of the photo--magnetic disintegration of the deuteron 
\begin{eqnarray}\label{label4.10}
{\cal M}({\gamma + D \to n + p}) =\frac{{\cal A}_{\rm
th}}{\displaystyle 1 - \frac{1}{2} r_{\rm np} a_{\rm np} k^2 + i\,a_{\rm
np} k}\, [\bar{u}(p_2)\gamma^5 u^c(p_1)],
\end{eqnarray}
where the factor $1/(1 - \frac{1}{2} r_{\rm np} a_{\rm np} k^2
+i\,a_{\rm np} k)$ describes the contribution of  low--energy elastic np
scattering in agreement with low--energy nuclear phenomenology
[17]. The amplitude Eq.(\ref{label4.10}) yields the cross section
\begin{eqnarray}\label{label4.11} 
\sigma^{\rm \gamma D}(\omega) = \sigma_0
\Bigg(\frac{\omega}{\varepsilon_{\rm D}}\Bigg) \frac{k r_{\rm
D}}{\displaystyle \Big(1 - \frac{1}{2} r_{\rm np} a_{\rm
np}k^2\Big)^2 + a^2_{\rm np} k^2}.
\end{eqnarray}
At zero effective range $r_{\rm np} = 0$ the cross section
Eq.(\ref{label4.11}) reduces to the form
\begin{eqnarray}\label{label4.12}
\sigma^{\rm \gamma D}(\omega) = \sigma_0
\Bigg(\frac{\omega}{\varepsilon_{\rm D}}\Bigg) \frac{k r_{\rm
D}}{\displaystyle 1 + a^2_{\rm np} k^2}.
\end{eqnarray}
In the PMA [17], in turn, at zero effective range the cross section
$\sigma^{\rm \gamma D}(\omega)$ has been found in the form
\begin{eqnarray}\label{label4.13}
\sigma^{\rm \gamma D}(\omega) = \sigma_0
\Bigg(\frac{\omega}{\varepsilon_{\rm D}}\Bigg)
\frac{k r_{\rm D}}{1 + a^2_{\rm np} k^2}\frac{1}{(1 +  r^2_{\rm D}k^2)^2}.
\end{eqnarray}
It is seen that Eqs.(\ref{label4.13}) and (\ref{label4.12}) differ by
a factor $1/(1 + r^2_{\rm D}k^2)^2$.  This factor as well as the
dependence on the S--wave scattering length and the effective range
[17] appears by virtue of the integral of the overlap of the wave
functions of the deuteron $\psi_{\rm D}(r)$ and the relative movement
of the np--system in the ${^1}{\rm S}_0$--state $\psi_{\rm np}(kr)$.

In order to introduce in the RFMD the wave functions of the deuteron
and the relative movement of the np--pair we can follow Bohr and
Mottelson [29] and: (1) in the initial nuclear state $|D(k_{\rm
D})\gamma(q)> = a^{\dagger}_{\rm D}(\vec{k}_{\rm D},\lambda_{\rm
D})a^{\dagger}(\vec{q},\lambda)|0>$ represent the operator of creation
of the deuteron $a^{\dagger}_{\rm D}(\vec{k}_{\rm D},\lambda_{\rm D})$
with 3--momentum $\vec{k}_{\rm D}$ and polarization $\lambda_{\rm D}$
in terms of the operators of creation of the proton $a^{\dagger}_{\rm
p}(\vec{p},\sigma_{\rm p})$ and the neutron $a^{\dagger}_{\rm
n}(\vec{k}_{\rm D} - \vec{p},\sigma_{\rm n})$ and (2) in the final
nuclear state $<n(p_2)p(p_1)| = <0|a_{\rm
n}(\vec{p}_2,\sigma_2)\,a_{\rm p}(\vec{p}_1,\sigma_1)$ replace the
product of the operators of annihilation of the neutron and the proton
by the operator of annihilation of the np--pair in the ${^1}{\rm
S}_0$--state $a_{\rm n}(\vec{p}_2,\sigma_2)\,a_{\rm
p}(\vec{p}_1,\sigma_1) \to a_{\rm np}(\vec{P},\vec{k}; S = 0)$, where
$\vec{P} = \vec{p}_1 + \vec{p}_2$, $\vec{k} = (\vec{p}_1 -
\vec{p}_2)/2$ and $S=0$ is a total spin.  In the form adjusted to our
problem these changes read
\begin{eqnarray}\label{label4.14}
&&a^{\dagger}_{\rm D}(\vec{k}_{\rm D},\lambda_{\rm D}) \sim \sum_{\sigma_{\rm
p},\sigma_{\rm n} = \pm 1} \int \frac{d^3p}{\displaystyle
\sqrt{2E_{\vec{p}}\,2E_{{\vec{k}_{\rm D} -
\vec{p}}}}}\,e_{\mu}(\vec{k}_{\rm D},\lambda_{\rm D})\,
[\bar{u}(\vec{p},\sigma_{\rm p})\gamma^{\mu} u^c(\vec{k}_{\rm D} -
\vec{p}, \sigma_{\rm n})]\,\nonumber\\ 
&&\times\,a^{\dagger}_{\rm
p}(\vec{p},\sigma_{\rm p})\,a^{\dagger}_{\rm n}(\vec{k}_{\rm D} -
\vec{p},\sigma_{\rm n}) \int d^3r\,\psi_{\rm
D}(r)\,e^{\textstyle i(\vec{p} - \vec{k}_{\rm D}/2)\cdot
\vec{r}},\nonumber\\ 
&&a_{\rm np}(\vec{P}, \vec{k}; S = 0) \sim
\sum_{\sigma_1,\sigma_2 = \pm 1} \int \frac{d^3p}{\displaystyle
\sqrt{2E_{\vec{p}}\,2E_{{\vec{P} -
\vec{p}}}}}\,[\bar{u^c}(\vec{p},\sigma_2)\gamma^5 u(\vec{P}-
\vec{p}, \sigma_1)]\,\nonumber\\ 
&&\times\,a_{\rm
n}(\vec{p},\sigma_2)\,a_{\rm p}(\vec{P} - \vec{p},\sigma_1) \int
d^3r\,\psi^*_{\rm np}(kr)\,e^{\textstyle - i(\vec{p} -
\vec{P}/2)\cdot \vec{r}},
\end{eqnarray}
where $E_{\vec{p}} = \sqrt{\vec{p}^{\,2} + M^2_{\rm N}}$.  The
spinorial parts of the wave functions of the deuteron and the np--pair
are given in terms of the Dirac bispinors in the
relativistically covariant form.  The operators Eq.(\ref{label4.14})
can be involved into evaluation of low--energy nuclear matrix elements
through the reduction technique [30].

However, such a modification complicates the model substantially and
goes beyond the scope of this paper. Therefore, referring to the
possibility to describing the factor $1/(1 + r^2_{\rm D}k^2)^2$
correctly in the RFMD the problem of this factor can be preferably
solved phenomenologically. In fact, since this factor is universal for
all processes of the deuteron coupled to the NN system in the
${^1}{\rm S}_0$--state at low energies, we suggest to multiply by a
factor $1/(1 + r^2_{\rm D}k^2)$ any amplitude of low--energy
nuclear process of this kind obtained near threshold, i.e. defined by
the corresponding effective Lagrangian evaluated through one--nucleon
loop exchanges in leading order in the large $N_C$ expansion [1]. This
implies the change
\begin{eqnarray}\label{label4.15}
{\cal A}_{\rm th} \to \frac{{\cal A}_{\rm th}}{1 + r^2_{\rm D}k^2}.
\end{eqnarray}
In other words, we introduce an universal form factor
\begin{eqnarray}\label{label4.16}
F_{\rm D}(k^2) =  \frac{1}{1 + r^2_{\rm D}k^2}
\end{eqnarray}
describing a spatial smearing of the deuteron coupled to the NN system
in the ${^1}{\rm S}_0$--state at low energies.

As a result  the amplitude of the photo--magnetic disintegration of
the deuteron obtained in the RFMD is equal to
\begin{eqnarray}\label{label4.17}
{\cal M}({\gamma + D \to n + p}) = \frac{{\cal A}_{\rm
th}}{\displaystyle 1 - \frac{1}{2}r_{\rm np}a_{\rm np}k^2 + i\,a_{\rm
np}\,k}\,F_{\rm D}(k^2)\, [\bar{u}(p_2)\gamma^5 u^c(p_1)].
\end{eqnarray}
The cross section for the photo--magnetic disintegration of the
 deuteron evaluated in the RFMD reads
\begin{eqnarray}\label{label4.18}
\sigma^{\rm \gamma D}(\omega) &=& \sigma_0
\Bigg(\frac{\omega}{\varepsilon_{\rm D}}\Bigg) \frac{k r_{\rm
D}}{\displaystyle \Big(1 - \frac{1}{2}r_{\rm np}a_{\rm np}k^2\Big)^2 +
a^2_{\rm np} k^2}\,F^2_{\rm D}(k^2)=\nonumber\\ 
&=& \sigma_0 \frac{k
r_{\rm D}}{1 + r^2_{\rm D}k^2}\,\frac{1}{\displaystyle \Big(1 -
\frac{1}{2}r_{\rm np}a_{\rm np}k^2\Big)^2 + a^2_{\rm np} k^2},
\end{eqnarray}
where $\sigma_0$ is given by Eq.(\ref{label4.2}).  

\section{Anti--neutrino disintegration of the deuteron via
charged weak current interaction}
\setcounter{equation}{0}

\hspace{0.2in} The effective Lagrangian describing the low--energy
nuclear transition $\bar{\nu}_{\rm e}$ + D $\to$ e$^+$ + n + n has
been calculated in [5,6] through one--nucleon loop exchanges and in
leading order in the large $N_C$ expansion [1]:
\begin{eqnarray}\label{label5.1}
{\cal L}_{\rm \bar{\nu}_{\rm e}D \to e^+ nn }(x) =-i g_{\rm A}M_{\rm N}
G_{\rm \pi
NN}\frac{G_{\rm V}}{\sqrt{2}}\frac{3g_{\rm
V}}{4\pi^2}\,D_{\mu}(x)\,[\bar{n}(x)\gamma^5
n^c(x)]\,[\bar{\psi}_{\nu_{\rm e}}(x)\gamma^{\mu}(1 - \gamma^5) \psi_{\rm
e}(x)],
\end{eqnarray}
where $G_{\rm V} = G_{\rm F}\,\cos \vartheta_C$ with $G_{\rm F} =
1.166\,\times\,10^{-11}\,{\rm MeV}^{-2}$ and $\vartheta_C$ are the
Fermi weak coupling constant and the Cabibbo angle $\cos \vartheta_C =
0.975$ and $g_{\rm A} = 1.267$; $\bar{\psi}_{\nu_{\rm e}}(x)$ and
$ \psi_{\rm e}(x)$ are the interpolating neutrino (anti--neutrino) and
electron (positron) fields.  The amplitude of $\bar{\nu}_{\rm e}$ + D
$\to$ e$^+$ + n + n we obtain in the form
\begin{eqnarray}\label{label5.2}
\hspace{-0.5in}i{\cal M}(\bar{\nu}_{\rm e} + {\rm D} \to {\rm e}^+ +
{\rm n} + {\rm n}) &=& -\,g_{\rm A} M_{\rm N}\,G_{\rm \pi
NN}\,\frac{G_{\rm V}}{\sqrt{2}}\,\frac{3g_{\rm
V}}{2\pi^2}\,\frac{\,F_{\rm D}(k^2)}{\displaystyle 1 
- \frac{1}{2}r_{\rm nn}a_{\rm
nn}k^2 + i\,a_{\rm nn}\,k}\, \nonumber\\ 
&&\times
e_{\mu}(Q)\,[\bar{v}(k_{\bar{\nu}_{\rm e}})\gamma^{\mu}(1-\gamma^5)
v(k_{{\rm e}^+})]\,[\bar{u}(p_1) \gamma^5 u^c(p_2)],
\end{eqnarray}
where the form factor $F_{\rm D}(k^2)$ provides a spatial smearing of
the deuteron.  The factor $1/(1 - \frac{1}{2}r_{\rm nn}a_{\rm nn}k^2 +
i\,a_{\rm nn}\,k)$ describes the nn interaction in the final state
which has been taken into account by summing up one--nucleon loop
diagrams, evaluated in leading order in the large $N_C$ expansion, and
renormalizing the wave functions of the neutrons. Since we work in the
isotopical limit, we set $a_{\rm nn} = a_{\rm np} = -
23.75\,{\rm fm}$ and $r_{\rm nn} = r_{\rm np} = 2.75\,{\rm fm}$. The
recent experimental values of the S--wave scattering length and the
effective range of low--energy elastic nn scattering are equal to
$a_{\rm nn}=( - 18.8\pm 0.3)\,{\rm fm}$ and $r_{\rm nn} = (2.75\pm
0.11)\,{\rm fm}$ [31,32].  

The amplitude Eq.\,(\ref{label5.2}),
squared, averaged over polarizations of the deuteron and summed over
polarizations of the final particles, reads
\begin{eqnarray}\label{label5.3}
\hspace{-0.2in}\overline{|{\cal M}(\bar{\nu}_{\rm e} + {\rm D} \to
{\rm e}^+ + {\rm n} + {\rm n})|^2} =\frac{144}{\pi^2}\,\frac{Q_{\rm
D}g^2_{\rm A}G^2_{\rm V}G^2_{\rm \pi NN}M^6_{\rm N}\,F^2_{\rm
D}(k^2)}{\displaystyle \Big(1 - \frac{1}{2}r_{\rm nn} a_{\rm nn}
k^2\Big)^2 + a^2_{\rm nn}k^2} \Big( E_{{\rm e}^+} E_{\bar{\nu}_{\rm
e}} - \frac{1}{3}\vec{k}_{{\rm e}^+}\cdot \vec{k}_{\bar{\nu}_{\rm
e}}\Big).
\end{eqnarray}
In the RFMD the momentum dependence of the amplitude of the
anti--neutrino disintegration of the deuteron agrees with that
obtained in the PMA [33]. A much more complicated momentum dependence
given in terms of the phenomenological form factors has been suggested
by Mintz [34].

The cross section for the process $\bar{\nu}_{\rm e}$ + D $\to$ e$^+$
+ n + n is defined by
\begin{eqnarray}\label{label5.4}
&&\sigma^{\bar{\nu}_{\rm e} D}_{\rm cc}(E_{\bar{\nu}_{\rm e}}) =
\frac{1}{4E_{\rm D}E_{\bar{\nu}_{\rm e}}}\int\,\overline{|{\cal
M}(\bar{\nu}_{\rm e} +
{\rm D} \to  {\rm e}^+ + {\rm n} + {\rm n})|^2}\nonumber\\
&&\frac{1}{2}\,(2\pi)^4\,\delta^{(4)}(Q + k_{{\bar{\nu}_{\rm e}}} - p_1 -
p_2 - k_{{\rm e}^+})\,
\frac{d^3p_1}{(2\pi)^3 2E_1}\frac{d^3 p_2}{(2\pi)^3 2E_2}\frac{d^3k_{{\rm
e}^+}}{(2\pi)^3
2E_{{\rm e}^+}},
\end{eqnarray}
where $E_{\rm D}$, $E_{\bar{\nu}_{\rm e}}$, $E_1$, $E_2$ and $E_{{\rm
e}^+}$ are the energies of the deuteron, the anti--neutrino, the
neutrons and the positron. The abbreviation (cc) denotes the charged
current. The integration over the phase volume of the (${\rm n n
e^+}$)--state we perform in the non--relativistic limit and in the
rest frame of the deuteron
\begin{eqnarray}\label{label5.5}
&&\frac{1}{2}\,\int\frac{d^3p_1}{(2\pi)^3 2E_1}\frac{d^3p_2}{(2\pi)^3
2E_2} \frac{d^3k_{{\rm e}^+}}{(2\pi)^3 2E_{{\rm
e}^+}}(2\pi)^4\,\delta^{(4)}(Q + k_{{\bar{\nu}_{\rm e}}} - p_1 - p_2 -
k_{{\rm e}^+})\,\nonumber\\ &&\times\, \frac{\displaystyle \Big(
E_{{\rm e}^+} E_{\bar{\nu}_{\rm e}} - \frac{1}{3} \vec{k}_{{\rm
e}^+}\cdot \vec{k}_{\bar{\nu}_{\rm e}}\Big)\,F^2_{\rm D}(M_{\rm N}\,T_{\rm
nn})}{\displaystyle \Big(1 -
\frac{1}{2} r_{\rm nn} a_{\rm nn} M_{\rm N}T_{\rm nn}\Big)^2 +
a^2_{\rm nn}M_{\rm N}\,T_{\rm nn}} = \frac{E_{\bar{\nu}_{\rm e}}M^3_{\rm
N}}{1024\pi^2}\,\Bigg(\frac{E_{\rm th}}{M_{\rm N}}
\Bigg)^{\!\!7/2}\Bigg(\frac{2 m_{\rm e}}{E_{\rm
th}}\Bigg)^{\!\!3/2}\frac{8}{\pi E^2_{\rm th}} \nonumber\\ &&\times
\int\!\!\!\int dT_{\rm e^+} dT_{\rm nn}\, \frac{\displaystyle
\sqrt{T_{\rm e^+}T_{\rm nn}}\,F^2_{\rm D}(M_{\rm N}\,T_{\rm
nn})}{\displaystyle \Big(1 -
\frac{1}{2} r_{\rm nn} a_{\rm nn} M_{\rm N}T_{\rm nn}\Big)^2 +
a^2_{\rm nn}M_{\rm N}\,T_{\rm nn}}\,\nonumber\\
&&\times\,\Bigg(1 + \frac{T_{\rm e^+}}{m_{\rm e}}
\Bigg)\,{\displaystyle \sqrt{1 + \frac{T_{\rm e^+}}{2 m_{\rm e}} }}\,
\delta\Big(E_{\bar{\nu}_{\rm e}}- E_{\rm th} - T_{\rm e^+} - T_{\rm
nn}\Big) = \nonumber\\ 
&&= \frac{E_{\bar{\nu}_{\rm e}}M^3_{\rm
N}}{1024\pi^2}\,\Bigg(\frac{E_{\rm th}}{M_{\rm N}}
\Bigg)^{\!\!7/2}\Bigg(\frac{2 m_{\rm e}}{E_{\rm
th}}\Bigg)^{\!\!3/2}\Bigg (\frac{E_{\bar{\nu}_{\rm e}}} {E_{\rm th}} -
1\Bigg)^{\!\!2}\,f\Bigg(\frac{E_{\bar{\nu}_{\rm e}}} {E_{\rm
th}}\Bigg),
\end{eqnarray}
where $T_{\rm nn}$ is the kinetic energy of the nn system, $T_{\rm
e^+}$ and $m_{\rm e} = 0.511\,{\rm MeV}$ are the kinetic energy and
the mass of the positron, $E_{\rm th}$ is the anti--neutrino energy
threshold of the reaction $\bar{\nu}_{\rm e}$ + D $\to$ e$^+$ + n +
n:  $E_{\rm th}= \varepsilon_{\rm D} + m_{\rm e} +
(M_{\rm n} - M_{\rm p}) = (2.225 + 0.511 + 1.293) \, {\rm MeV} =
4.029\,{\rm MeV}$. The function $f(y)$, where $y=E_{\bar{\nu}_{\rm
e}}/E_{\rm th}$, is defined as
\begin{eqnarray}\label{label5.6}
f(y) &=& \frac{8}{\pi}\,\int\limits^{1}_{0} dx\, \frac{\sqrt{x\,(1 -
x)}\,F^2_{\rm D}(M_{\rm N}E_{\rm th}\,(y - 1)\,x)}{\displaystyle
\Big(1 - \frac{1}{2}r_{\rm nn} a_{\rm nn}M_{\rm N}E_{\rm th}\,(y -
1)\,x \Big)^2 + a^2_{\rm nn}M_{\rm N}E_{\rm th}\,(y -
1)\,x}\nonumber\\ &&\times\,\Big(1 + \frac{E_{\rm th}}{m_{\rm
e}}(y-1)(1-x)\Big) \,\sqrt{1 + \frac{E_{\rm th}} {2 m_{\rm
e}}(y-1)(1-x)},
\end{eqnarray}
where we have changed the variable $T_{\rm nn} = (E_{\bar{\nu}_{\rm e}} -
E_{\rm th})\,x$.
The function $f(y)$ is normalized to unity at $y=1$, i.e.  at threshold
$E_{\bar{\nu}_{\rm e}} = E_{\rm th}$. Thus, the cross section for the
anti--neutrino disintegration
of the deuteron reads
\begin{eqnarray}\label{label5.7}
\sigma^{\rm \bar{\nu}_{\rm e} D}_{\rm cc}(E_{\bar{\nu}_{\rm e}}) = \sigma_0\,(y -
1)^2\,f(y),
\end{eqnarray}
where $\sigma_0$ is given by
\begin{eqnarray}\label{label5.8}
\hspace{-0.5in}\sigma_0 = Q_{\rm D}\,G^2_{\rm \pi NN}\,\frac{9g^2_{\rm A}
G^2_{\rm V} M^8_{\rm N}}{512\pi^4}\,\Bigg(\frac{E_{\rm th}}{M_{\rm
N}}\Bigg)^{\!\!7/2}\Bigg(\frac{2 m_{\rm e}}{E_{\rm th}}\Bigg)^{\!\!3/2}
= 4.58\,\times \,10^{-43}\,{\rm cm}^2.
\end{eqnarray}
 The value $\sigma_0 = 4.58\,\times\,10^{-43} {\rm cm}^2$ agrees with
the value $\sigma_0 = 4.68\,\times \,10^{-43}\,{\rm cm}^2$ obtained in
the PMA [33] (see Fig.\,7 of Ref.\,[13]).

The experimental data on the anti--neutrino disintegration of the
deuteron are given in terms of the cross section averaged over the
anti--neutrino energy spectrum [14]:\\
 $<\sigma^{\rm \bar{\nu}_{\rm e}
D}_{\rm cc}(E_{\bar{\nu}_{\rm e}})>_{\exp} = (9.83\pm 2.04)\times
10^{-45}\,{\rm cm}^2$.

In order to average the theoretical cross section Eq.(\ref{label5.7})
over the anti--neutrino spectrum we should use the spectrum given by
Table YII of Ref.[14].  This yields
\begin{eqnarray}\label{label5.9}
<\sigma^{\rm \bar{\nu}_{\rm e} D}_{\rm cc}(E_{\bar{\nu}_{\rm e}})> = 
11.7\times 10^{-45}\,{\rm cm}^2.
\end{eqnarray}
The theoretical value Eq.\,(\ref{label5.9}) agrees good with the
experimental one $<\sigma^{\rm \bar{\nu}_{\rm e} D}_{\rm
cc}(E_{\bar{\nu}_{\rm e}})>_{\exp} = (9.83\pm 2.04)\times
10^{-45}\,{\rm cm}^2$ [14].

\section{Neutrino and anti--neutrino disintegration of the deuteron via
neutral weak current interaction} 
\setcounter{equation}{0}

\hspace{0.2in} The amplitude of the neutrino disintegration of
the deuteron caused by neutral weak current $\nu_{\rm e}$ + D
$\to$ $\nu_{\rm e}$ + n + p can be evaluated by analogy with the
amplitude of the reaction $\bar{\nu}_{\rm e}$ + D $\to$ e$^+$ + n + n
through one--nucleon loop exchanges (see Ref.[6]) and in leading order
in the large $N_C$ expansion [1]:
\begin{eqnarray}\label{label6.1}
&&i{\cal M}(\nu_{\rm e} + {\rm D} \to \nu_{\rm e} + {\rm n} + {\rm p})
= - g_{\rm A}\, M_{\rm N}\,\frac{G_{\rm F}}{\sqrt{2}}\,\frac{3g_{\rm
V}}{4\pi^2}\, \frac{G_{\rm \pi NN}\,F_{\rm D}(k^2)}{\displaystyle 1 
- \frac{1}{2}r_{\rm np}a_{\rm
np}k^2 + i\,a_{\rm np}\,k}\,\nonumber\\ &&\times\,e_{\mu}(k_{\rm
D})\,[\bar{u}(k^{\prime}_{\nu_{\rm e}})\gamma^{\mu}(1-\gamma^5)
u(k_{\nu_{\rm e}})] [\bar{u}(p_1) \gamma^5
u^c(p_2)],
\end{eqnarray}
where $\bar{u}(k^{\prime}_{\nu_{\rm e}})$, $u(k_{\nu_{\rm e}})$,
$\bar{u}(p_1)$ and $u^c(p_2)$ are the Dirac bispinors of the initial
and the final neutrinos, and the nucleons. Then, the form factor
$F_{\rm D}(k^2)$ provides a spatial smearing of the deuteron and the
factor $1/(1 - \frac{1}{2}r_{\rm np}a_{\rm np}k^2 + i\,a_{\rm np}\,k)$
describes the np interaction in the final state.

The amplitude Eq.(\ref{label6.1}) squared, averaged over
polarizations of the deuteron, summed over polarizations of the
nucleons reads
\begin{eqnarray}\label{label6.2}
\overline{|{\cal M}(\nu_{\rm e} + {\rm D} \to \nu_{\rm e} + {\rm n}
+{\rm p})|^2} = \frac{36
}{\pi^2}\,
\frac{Q_{\rm D} g^2_{\rm A}G^2_{\rm F}G^2_{\rm \pi NN}
M^6_{\rm N}F^2_{\rm D}(k^2)}{\displaystyle \Big(1 -
\frac{1}{2} r_{\rm np} a_{\rm np}  k^2 \Big)^2 + a^2_{\rm np}k^2
}\Big( E^{\prime}_{\nu_{\rm e}} E_{\nu_{\rm e}} - \frac{1}{3}
\vec{k}^{\prime}_{\nu_{\rm e}} \cdot \vec{k}_{\nu_{\rm e}}\Big).
\end{eqnarray}
In the rest frame of the deuteron the cross section for the process
$\nu_{\rm e}$ + D $\to$ $\nu_{\rm e}$ + n + p is defined
\begin{eqnarray}\label{label6.3}
&&\sigma^{\nu_{\rm e} D}_{\rm nc}(E_{\bar{\nu}_{\rm e}}) =
\frac{1}{4M_{\rm D}E_{\nu_{\rm e}}}\int\,\overline{|{\cal
M}(\nu_{\rm e} +
{\rm D} \to \nu_{\rm e} + {\rm n} + {\rm p})|^2}\nonumber\\
&&(2\pi)^4\,\delta^{(4)}(k_{\rm D} + k_{\nu_{\rm e}} - p_1 -
p_2 - k^{\prime}_{\nu_{\rm e}})\,
\frac{d^3p_1}{(2\pi)^3 2E_1}\frac{d^3 p_2}{(2\pi)^3
2E_2}\frac{d^3k^{\prime}_{\nu_{\rm e}}}{(2\pi)^3
2E^{\prime}_{\nu_{\rm e}}}.
\end{eqnarray}
The abbreviation (nc) denotes the neutral current. The integration
over the phase volume of the (${\rm n p \nu_{\rm e}}$)--state we
perform in the non--relativistic limit and in the rest frame of the
deuteron,
\begin{eqnarray}\label{label6.4}
&&\int\frac{d^3p_1}{(2\pi)^3 2E_1}\frac{d^3p_2}{(2\pi)^3 2E_2}
\frac{d^3k^{\prime}_{\nu_{\rm e}}}{(2\pi)^3 2E^{\prime}_{\nu_{\rm
e}}}(2\pi)^4\,\delta^{(4)}(k_{\rm D} + k_{\nu_{\rm e}} - p_1 - p_2 -
k^{\prime}_{\nu_{\rm e}})\,\nonumber\\
&&\Big( E_{\nu_{\rm e}} E^{\prime}_{\nu_{\rm e}} - \frac{1}{3}
\vec{k}_{\nu_{\rm e}}\cdot
\vec{k}^{\prime}_{\nu_{\rm e}}\Big)\,\frac{F^2_{\rm D}(M_{\rm N}T_{\rm
np})}{\displaystyle \Big(1 - \frac{1}{2}r_{\rm np}a_{\rm np} M_{\rm
N}T_{\rm np} \Big)^2 + a^2_{\rm np}M_{\rm N}T_{\rm np}} = \nonumber\\
&&=\frac{E_{\nu_{\rm e}}M^3_{\rm N}}{210\pi^3}\,\Bigg(\frac{E_{\rm
th}}{M_{\rm N}}\Bigg)^{\!\!7/2}\,(y - 1)^{7/2}\,\Omega_{\rm np\nu_{\rm e}}(y).
\end{eqnarray}
The function $\Omega_{\rm np\nu_{\rm e}}(y)$, where
$y=E_{\bar{\nu}_{\rm e}}/E_{\rm th}$ and $E_{\rm th}= \varepsilon_{\rm
D}=2.225\,{\rm MeV}$ is threshold of the reaction, is defined as
\begin{eqnarray}\label{label6.5}
\Omega_{\rm np\nu_{\rm e}}(y) = \frac{105}{16}\,\int\limits^{1}_{0} dx
\frac{\sqrt{x}\,(1 - x)^2}{\displaystyle \Big(1 - \frac{1}{2}\frac{r_{\rm np}a_{\rm
np} }{r^2_{\rm D}}(y-1) x \Big)^2 + \frac{a^2_{\rm np}}{r^2_{\rm
D}}(y-1)\, x}\frac{1}{(1 + (y - 1)\, x)^2},
\end{eqnarray}
where we have changed the variable $T_{\rm np} = (E_{\bar{\nu}_{\rm
e}} - E_{\rm th})\,x$ and used the relation $M_{\rm N}E_{\rm th} =
1/r^2_{\rm D}$ at $E_{\rm th}=\varepsilon_{\rm D}$. The function
$\Omega_{\rm np\nu_{\rm e}}(y)$ is normalized to unity at $y=1$, i.e.,
at threshold $E_{\bar{\nu}_{\rm e}} = E_{\rm th}$.

The cross section for the neutrino disintegration of the deuteron
caused by the neutral weak current $\nu_{\rm e}$ + D $\to$ $\nu_{\rm
e}$ + n + p reads
\begin{eqnarray}\label{label6.6}
\sigma^{\rm \nu_{\rm e} D}_{\rm nc}(E_{\nu_{\rm e}}) = \sigma_0\,(y -
1)^{7/2}\,\Omega_{\rm np\nu_{\rm e}}(y),
\end{eqnarray}
where $\sigma_0$ is defined by
\begin{eqnarray}\label{label6.7}
\sigma_0 = Q_{\rm D}\,G^2_{\rm \pi NN}\,\frac{3 g^2_{\rm A}\,G^2_{\rm
F}  M^8_{\rm N}}{140\pi^5}\,
\Bigg(\frac{E_{\rm th}}{M_{\rm N}}\Bigg)^{\!\!7/2}= 1.84 \times
10^{-43}\,{\rm cm}^2.
\end{eqnarray}
In our approach the cross section for the disintegration of the
deuteron by neutrinos $\nu_{\rm e}$ + D $\to$ $\nu_{\rm e}$ + n + p
coincides with the cross section for the disintegration of the
deuteron by anti--neutrinos $\bar{\nu}_{\rm e}$ + D $\to$
$\bar{\nu}_{\rm e}$ + n + p, $\sigma^{\rm \nu_{\rm e}D}_{\rm
nc}(E_{\nu_{\rm e}}) = \sigma^{\rm \bar{\nu}_{\rm e}D}_{\rm
nc}(E_{\bar{\nu}_{\rm e}})$. Therefore, we compare our results with
the experimental data on the disintegration of the deuteron by
anti--neutrinos [14]. The experimental value of the cross section for
the anti--neutrino disintegration of the deuteron $\bar{\nu}_{\rm e}$
+ D $\to$ $\bar{\nu}_{\rm e}$ + n + p averaged over the anti--neutrino
spectrum reads [14]: $<\sigma^{\rm \bar{\nu}_{\rm e}D}_{\rm
nc}(E_{\bar{\nu}_{\rm e}})>_{\exp} = (6.08\pm 0.77)\times
10^{-45}\,{\rm cm}^2$.

By using the anti--neutrino spectrum given
by Table YII of Ref.[14] for the calculation of the average value
of the theoretical cross section Eq.(\ref{label6.6}) we obtain
\begin{eqnarray}\label{label6.8}
<\sigma^{\rm \bar{\nu}_{\rm e} D}_{\rm nc}(E_{\bar{\nu}_{\rm e}})> = 
6.4\times 10^{-45}\,{\rm cm}^2.
\end{eqnarray}
The theoretical value Eq.\,(\ref{label6.8}) agrees good with the
experimental one $<\sigma^{\rm \bar{\nu}_{\rm e}D}_{\rm
nc}(E_{\bar{\nu}_{\rm e}})>_{\exp} = (6.08\pm 0.77)\times
10^{-45}\,{\rm cm}^2$ [14].

\section{Conclusion}
\setcounter{equation}{0}

\hspace{0.2in} The main goal of the paper is to show that: 1) Chiral
perturbation theory can be incorporated in the RFMD and 2) the
amplitudes of low--energy elastic nucleon--nucleon scattering
contributing to the reactions of the photo--magnetic and
anti--neutrino disintegration of the deuteron can be described in the
RFMD in agreement with low--energy nuclear phenomenology.

By example of the evaluation of the amplitude for the radiative
M1--capture n + p $\to$ D + $\gamma$ we have shown that Chiral
perturbation theory can be incorporated into the RFMD. We have
considered chiral meson--loop corrections from the virtual meson
transitions $\pi \to a_1 \gamma$, $a_1 \to \pi\,\gamma$, $\pi \to
(\omega, \rho) \gamma$, $(\omega, \rho) \to \pi \gamma$, $\sigma \to
(\omega, \rho) \gamma$ and $(\omega, \rho) \to \sigma \gamma$, where
$\sigma$ is a scalar partner of pions under chiral $SU(2)\times SU(2)$
transformations in (CHPT)$_q$ with a linear realization of chiral
$U(3)\times U(3)$ symmetry.  These virtual meson transitions give
contributions to the effective interactions of nucleons $\delta {\cal
L}^{\rm NN \gamma}_{\rm eff}(x)$ coupled to a magnetic field
\begin{eqnarray}\label{label7.1}
\delta {\cal L}^{\rm NN\gamma}_{\rm eff}(x) &=& \frac{ie}{4M_{\rm
N}}\,\mu^{(\chi)}_{\rm N}\, \bar{N}(x)\sigma_{\mu\nu}N(x)
F^{\mu\nu}(x).
\end{eqnarray}
The effective magnetic moments $\mu^{(\chi)}_{\rm N}$, caused by
chiral meson--loop corrections, have been evaluated in leading order
in the large $N_C$ expansion [1] and renormalized according to the
renormalization procedure developed in (CHPT)$_q$ for the evaluation
of chiral meson--loop corrections (see {\it Ivanov} in
Refs. [9]). Since the renormalized expressions should vanish in the
chiral limit $M_{\pi} \to 0$, the contributions of the virtual meson
transitions with intermediate $\sigma$--meson, finite in the chiral
limit, have been subtracted [9]. Such a cancellation of the
$\sigma$--meson contributions in the one--meson loop approximation
agrees with Chiral perturbation theory using a non--linear realization
of chiral symmetry, where $\sigma$--meson like exchanges can appear
only in two--meson loop approximation. The non--trivial contributions
vanishing in the chiral limit have been obtained only from the virtual
meson transitions with intermediate $\pi$--meson.

The numerical value of the cross section for the M1--capture
accounting for the contributions of chiral one--meson loop corrections
amounts to $\sigma^{\rm np}(k) = 287.2\,{\rm m b}$. This value differs
from the experimental one $\sigma^{\rm np}(k)_{\exp} = (334.2\pm
0.5)\,{\rm m b}$ by about 14$\%$. For the description of this
discrepancy we have taken into account the contribution of the
$\Delta(1232)$ resonance. The total cross section for the
neutron--proton radiative capture has been found dependent on the
parameter $Z$ defining the ${\rm \pi\Delta N}$ coupling off--mass
shell of the $\Delta(1232)$ resonance: $\sigma^{\rm n p}(k)= 287.2\,(1
+ 0.64\,(1-2Z))^2\,{\rm m\,b}$ Eq.(\ref{label3.23}). In order to fit
the experimental value of the cross section we should set
$Z=0.438$. This agrees with the experimental bound $|Z|\le 1/2$ [23].

When matching our result for the cross section for the M1--capture
with the recent one obtained in the EFT approach by Chen, Rupak and
Savage [35]: $\sigma^{\rm np}(k) =( 287.1 + 6.51 {^{\not\pi}}L_1)\,{
\rm m b}$ (see Eq.(3.49) of Ref.\,[35]), we accentuate the dependence
of the cross section on the parameter ${^{\not\pi}}L_1$ which has been
fixed from the experimental data. Since the contribution of the
$\Delta(1232)$ resonance has not been considered in Ref.[35] and there
is no uncertainties related to the parameter $Z$, ${^{\not\pi}}L_1$ is
a free parameter of the approach. Unlike Ref.[35] the cross section
calculated in the RFMD does not contain free parameters.

The cross section for the photo--magnetic disintegration of the
deuteron $\gamma$ + D $\to$ n + p has been evaluated for energies far
from threshold. For this aim we have taken into account the np
interaction in the final state by summing up an infinite series of
one--nucleon loop diagrams which have been calculated in leading order
in the large $N_C$ expansion. This has given the amplitude of
low--energy elastic np scattering contributing to the amplitude of
$\gamma$ + D $\to$ n + p in the form agreeing with low--energy nuclear
phenomenology, i.e. defined by the S--wave scattering length $a_{\rm
np}$ and the effective range $r_{\rm np}$. This result relaxes
substantially the statement by Bahcall and Kamionkowski [18] that in
the RFMD due to the effective local four--nucleon interaction
Eq.(\ref{label1.1}) one cannot describe low--energy elastic NN
scattering in agreement with low--energy nuclear
phenomenology. Nevertheless, the problem of the description of
low--energy elastic pp scattering accounting for the Coulomb repulsion
still remains.

We have shown that the dependence of the amplitude of the
photo--magnetic disintegration of the deuteron on the deuteron radius
$r_{\rm D}$ in the from of the factor $1/(1 + r^2_{\rm D}k^2)$ can be
justified in the RFMD by means of the direct inclusion of the wave
functions of the deuteron and the np--pair in the ${^1}{\rm
S}_0$--state.  However, such an inclusion leads to significant
complexification of the model consideration of which goes beyond
the scope of this paper.  The problem of the factor $1/(1 + r^2_{\rm
D}k^2)$, universal for all low--energy processes of the deuteron
coupled to the NN system in the ${^1}{\rm S}_0$--state, can be
preferably solved phenomenologically. Referring to the possibility to
derive this factor in the RFMD by the way having been discussed in
Sect.\,3 we have suggested to multiply the amplitudes of low--energy
nuclear transitions evaluated near thresholds by a factor $1/(1 +
r^2_{\rm D}k^2)$. In other words, we have introduced an universal form
factor $F_{\rm D}(k^2) = 1/(1 + r^2_{\rm D}k^2)$ describing a spatial
smearing of the deuteron coupled to the NN system in the the ${^1}{\rm
S}_0$--state.

This procedure has been applied to the evaluation of the cross
sections for the anti--neutrino disintegration of the deuteron caused
by charged $\bar{\nu}_{\rm e}$ + D $\to$ e$^+$ + n + n
and neutral $\bar{\nu}_{\rm e}$ + D $\to$ $\bar{\nu}_{\rm
e}$ + n + p weak currents. The theoretical cross sections averaged over the
anti--neutrino spectrum $<\sigma^{\rm \bar{\nu}_{\rm e} D}_{\rm
cc}(E_{\bar{\nu}_{\rm e}})> = 11.7\times 10^{-45}\,{\rm cm}^2$ and
$<\sigma^{\rm \bar{\nu}_{\rm e} D}_{\rm nc}(E_{\bar{\nu}_{\rm e}})> =
6.4\times 10^{-45}\,{\rm cm}^2$ agree good with recent experimental
data $<\sigma^{\bar{\nu}_{\rm e}D}_{\rm cc}(E_{\bar{\nu}_{\rm
e}})>_{\exp} = (9.83\pm 2.04) \times 10^{-45}\,{\rm cm}^2$ and
$<\sigma^{\bar{\nu}_{\rm e}D}_{\rm nc}(E_{\bar{\nu}_{\rm e}})>_{\exp}
= (6.08\pm 0.77) \times 10^{-45}\,{\rm cm}^2$ obtained  by the Reines's 
experimental group [14].

The cross sections for the reactions $\bar{\nu}_{\rm e}$ + D $\to$
e$^+$ + n + n and $\bar{\nu}_{\rm e}$ + D $\to$ $\bar{\nu}_{\rm e}$ +
n + p have been recently calculated by Butler and Chen [36] in the EFT
approach. The obtained results have been written in the
following general form $\sigma = (a + b\,L_{1,\rm A})\times
10^{-42}\,{\rm cm}^2$ (see Table I of Ref.\,[36]), where $a$ and $b$ are
the parameters which have been calculated in the approach, whereas
$L_{1,\rm A}$ is a free one. Thus, unlike the cross sections given by
Eqs.(\ref{label5.7}) and (\ref{label6.6}), where there are no free
parameters, the cross sections for the anti--neutrino disintegration
of the deuteron [36] as well as for the neutron--proton radiative
capture [35] calculated in the EFT approach depend on free
parameters. Due to independence of the cross sections
Eqs.(\ref{label5.7}) and (\ref{label6.6}) on free parameters we can
analyse and value in the RFMD not only chiral meson--loop corrections
but the corrections mentioned recently by Vogel and Beacom [37].

\section*{Acknowledgement}

We are grateful to Prof. Kamionkowski for helpful remarks and
encouragement for further applications of the expounded in the paper
technique to the evaluation of the astrophysical factor for pp fusion
and the cross section for the neutrino disintegration of the deuteron
$\nu_{\rm e}$ + D $\to$ e$^-$ + p + p by accounting for the Coulomb
repulsion between the protons. 

We thank Prof. Beacom for calling our attention to the experimental
data [14]. Discussions of the experimental data with Prof. Sobel
and Prof. Price are greatly appreciated.


\newpage

\begin{thebibliography}{9}
\bibitem{[1]}
 A. N. Ivanov, H. Oberhummer, N. I. Troitskaya and
M. Faber, 
{\it The relativistic field theory model of the deuteron
from low--energy QCD}, nucl--th/9908029, August 1999.
\bibitem{[2]}
A. N. Ivanov, N. I. Troitskaya, M. Faber and H. Oberhummer,
Phys. Lett. B361 (1995) 74.
\bibitem{[3]}
A. N. Ivanov, N. I. Troitskaya, M. Faber and H. Oberhummer,
Nucl. Phys. A617 (1997) 414 and references therein.
\bibitem{[4]}
A. N. Ivanov, N. I. Troitskaya, M. Faber and H. Oberhummer,
Nucl. Phys. A625 (1997) 896 (Erratum).
\bibitem{[5]} 
A. N. Ivanov, H. Oberhummer, N. I. Troitskaya and
M. Faber, 
{\it Solar proton burning, photon and anti--neutrino
disintegration of the deuteron in the relativistic field theory model
of the deuteron}, nucl--th/9810065, October 1998.
\bibitem{[6]} 
A. N. Ivanov, H. Oberhummer, N. I. Troitskaya and
M. Faber, 
{\it Solar neutrino processes in the relativistic field theory model 
of the deuteron}, nucl--th/9811012, November 1998.
\bibitem{[7]}
A. E. Cox, A. R. Wynchank and C. H. Collie, 
Nucl. Phys. 74 (1965) 497 and references therein.
\bibitem{[8]} 
T.--S. Park, D.--P. Min and M. Rho, 
Phys. Rev. Lett. 74
(1995) 4153; Nucl. Phys. A596 (1996) 515;
T.--S. Park, K. Kubodera, D.--P. Min and M. Rho, 
{\it Effective field theory approach to {\rm
n}(polarized) {\rm + p}(polarized) $\to$ {\rm D +} $\gamma$  at
threshold}, nucl--th/9906005, June 1999.
\bibitem{[9]} 
A. N. Ivanov, M. Nagy and N. I. Troitskaya,
Int. J. Mod. Phys. A7 (1992) 7305; 
A. N. Ivanov, Int. J. Mod. Phys. A8
(1993) 853; 
A. N. Ivanov, N. I. Troitskaya and M. Nagy,
Int. J. Mod. Phys. A8 (1993) 2027; {\it ibid.} A8 (1993) 3425;
Phys. Lett. B308 (1993) 111; {\it ibid.} B295 (1992) 308; 
A. N. Ivanov and N. I. Troitskaya, 
Nuovo Cim. A108 (1995) 555; Phys. Lett. B342
(1995) 323; {\it ibid.} B387 (1996) 386; {\it ibid.} B388 (1996) 869 
(Erratum); {\it ibid.} B390 (1997) 341; 
F. Hussain, A. N. Ivanov and N. I. Troitskaya, 
Phys. Lett. B348 (1995) 609; {\it ibid.} B369 (1996)
351.
\bibitem{[10]}
A. N. Ivanov, M. Nagy and N. I. Troitskaya,
Phys. Rev. C59 (1999) 451;
Ya. A. Berdnikov, A. N. Ivanov, V. F. Kosmach and N. I. Troitskaya,
Phys. Rev. C60 (1999) 015201.
\bibitem{[11]}
D. O. Riska and G. E. Brown,
Phys. Lett. B38 (1972) 193.
\bibitem{[12]}
A. N. Ivanov, N. I. Troitskaya, H. Oberhummer and M. Faber,
Z. Phys. A358 (1997) 81.
\bibitem{[13]}
T. L. Jenkins, F. E. Kinard and F. Reines,
Phys. Rev. 185 (1969) 1599.
\bibitem{[14]}
S. P. Riley, Z. D. Greenwood, W. R. Kroop, L. R. Price, 
F. Reines, H. W. Sobel, Y. Declais, A. Etenko and M. Skorokhvatov,
Phys. Rev. C59 (1999) 1780. 
\bibitem{[15]}
S. Weinberg,
Phys. Lett. B251 (1990) 288;
Nucl. Phys. B363 (1991) 3;
Phys. Lett. B295 (1992) 114.
\bibitem{[16]}
D. R. Kaplan, M. J. Savage and M. B. Wise,
Nucl. Phys. B478 (1996) 629 and references therein;
S. R. Beane, T. D. Cohen and D. R. Phillips,
Nucl. Phys. A632 (1998) 445;
T.--S. Park, K. Kubodera, D.--P. Min and M. Rho,
Nucl. Phys. A646 (1999) 83.
\bibitem{[17]} 
S. DeBenedetti, 
in {\it NUCLEAR INTERACTIONS}, John
Wiley $\&$ Sons, Inc., New York, 1964 and references therein.
\bibitem{[18]}
J. N. Bahcall and M. Kamionkowski,
Nucl. Phys. A625 (1997) 893.
\bibitem{[19]} 
C. Caso {\it et al.},
Eur. Phys. J.  C3 (1998) 1.
\bibitem{[20]}
W. Rarita and J. Schwinger,
Phys. Rev. 60 (1941) 61.
\bibitem{[21]}
L. M. Nath, B. Etemadi and J. D. Kimel,
Phys. Rev. D3 (1971) 2153.
\bibitem{[22]}
J. Kambor, 
{\it The $\Delta(1232)$ as an Effective Degree of Freedom in Chiral Perturbation Theory},
 Talk given at the Workshop on Chiral Dynamics, 1997 Mainz, Germany, September 1--5, 1997; hep--ph/9711484 26 November 1997.
\bibitem{[23]} K. Kabir, T. K. Dutta, Muslema Pervin and L. M. Nath,
{\it The Role of $\Delta(1232)$ in Two--pion Exchange Three--nucleon
Potential}, hep--th/9910043, October 1999.
\bibitem{[24]}
V. Bernard, N. Kaiser and Ulf--G. Meissner,
Int. J. Mod. Phys. E4 (1995) 193 and references therein.
\bibitem{[25]}
R. D. Peccei,
Phys. Rev. 181 (1969) 1902 and references therein.
\bibitem{[26]}
M. G. Olsson and E. T. Osypowski,
Nucl.Phys. B87 (1975) 399.
\bibitem{[27]}
M. M. Nagels et al.,
Nucl. Phys. B147 (1979) 253.
\bibitem{[28]}
I. S. Gertsein and R. Jackiw,
Phys. Rev. 181 (1969) 1955.
\bibitem{[29]} 
A. Bohr and B. R. Mottelson, in {\it NUCLEAR
STRUCTURE}, Vol. I, W. A. Benjamin, Inc., New York, 1969.
\bibitem{[30]} 
C. Itzykson and J.--B. Zuber, 
in {\it QUANTUM FIELD THEORY}, 
McGraw--Hill Book Company, New York, 1980, p. 217.
\bibitem{[31]} 
G. A. Miller, B. M. K. Nefkens and I. Slaus,
Phys. Rep. 194 (1990) 1; 
G. A. Miller and W. N. T. van Oers, 
in {\it Symmetries and Fundamental Interactions in Nuclei}, 
W. C. Haxton and E. M. Henley, eds.,World Scientific Singapore, 
1995, p.127.
\bibitem{[32]} 
R. Machchleidt and M. K. Banerjee, 
{\it Charge--dependence of the ${\rm \pi NN}$ coupling constant and
charge--dependence of the ${\rm NN}$ interaction}, nucl--th/9908066,
August 1999.
\bibitem{[33]}
J. Weneser,
Phys. Rev. 105 (1957) 1335;
T. Ahrens and T. P. Lang,
Phys. Rev. C3 (1971) 979;
J. Ho$\check{\rm s}$ek and E. Truhlik,
Phys. Rev. C23 (1981) 665;
F. T. Avignone,
Phys. Rev. D24 (1981) 778.
\bibitem{[34]}
S. L. Mintz,
Phys. Rev. C23 (1981) 421, ibid. C24 (1981) 1799.
\bibitem{[35]} 
J.--W. Chen, G. Rupak and M. J. Savage,
Nucl. Phys. A653 (1999) 386.
\bibitem{[36]} 
M. Butler and J.--W. Chen, {\it Elastic and Inelastic
neutrino--Deuteron Scattering in Effective Field Theory},
nucl--th/9905059, June 1999.
\bibitem{[37]} 
P. Vogel and J. F. Beacom,
Phys. Rev. D60 (1999) 053003.
\end{thebibliography}
\end{document}